\documentclass[journal,dvipsnames]{IEEEtran}

\usepackage[noadjust]{cite}
\ifCLASSINFOpdf
   \usepackage[pdftex]{graphicx}

\else
  \usepackage[dvips]{graphicx}
\fi
\usepackage{amsmath,amssymb,amsthm}
\usepackage{float}
\usepackage{color}
\usepackage{algorithmic}
\usepackage{array}
\ifCLASSOPTIONcompsoc
\usepackage[caption=false,font=normalsize,labelfont=sf,textfont=sf]{subfig}
\else
  \usepackage[caption=false,font=footnotesize]{subfig}
\fi
\usepackage{fixltx2e}
\usepackage{stfloats}
\usepackage{url}
\usepackage[hidelinks]{hyperref}
\usepackage{tikz}
\usepackage{pgfplots}
\usepackage{tikzscale}
\usetikzlibrary{shadows}
\usetikzlibrary{positioning}
\usepackage{balance}
\usepackage{epstopdf}
\usepackage{psfrag,pst-pdf}
\usepackage{orcidlink}
\usepackage{caption}

\usepackage{acronym}
\acrodef{2D}{2-dimensional}
\acrodef{3D}{3-dimensional}
\acrodef{3GPP}{3rd generation partnership project}
\acrodef{4G}{4th generation}
\acrodef{5G}{5th generation}
\acrodef{6G}{6th generation}
\acrodef{AC}{alternating current}
\acrodef{ADT}{angular diversity transmitter}
\acrodef{ADR}{angular diversity receiver}
\acrodef{AI}{artificial intelligence}
\acrodef{ANN}{artificial neural network}
\acrodef{ANSI}{American national standards institute}
\acrodef{AoA}{angle-of-arrival}
\acrodef{AoD}{angle-of-departure}
\acrodef{AP}{access point}
\acrodef{APD}{avalanche photodiode}
\acrodef{AR}{augmented reality}
\acrodef{AWG}{arbitrary waveform generator}
\acrodef{AWGN}{additive white Gaussian noise}
\acrodef{AWGR}{high port count arrayed waveguide grating router}
\acrodef{BER}{bit error ratio}
\acrodef{BPF}{bandpass filter}
\acrodef{BSM}{basic safety message}
\acrodef{C-V2X}{cellular-based vehicular communication}
\acrodef{CFAR}{constant false alarm rate}
\acrodef{CAM}{cooperative awareness message}
\acrodef{CCD}{charge-coupled device}
\acrodef{CDF}{cumulative distribution function}
\acrodef{CF}{consumption factor}
\acrodef{CIR}{channel impulse response}
\acrodef{cm}{centimeter}
\acrodef{CMS}{continuous mode selection}
\acrodef{CNN}{convolutional neural networks}
\acrodef{CSI}{channel state information}
\acrodef{CPC}{compound parabolic concentrator}
\acrodef{CoMB}{coordinated multi-beam}
\acrodef{CoMP}{coordinated multipoint}
\acrodef{DBC}{dynamic beam clustering}
\acrodef{DC}{direct current}
\acrodef{DCO{-}OFDM}{direct current-biased optical \ac{OFDM}}
\acrodef{DCN}{data center network}
\acrodef{DENM}{decentralized environmental notification message}
\acrodef{DFB}{distributed feedback laser}
\acrodef{DNN}{deep neural network}
\acrodef{DSRC}{dedicated short-range communication}
\acrodef{EGC}{equal gain combining}
\acrodef{EMI}{electromagnetic interference}
\acrodef{ES}{exhaustive search}
\acrodef{FCNN}{fully connected neural network}
\acrodef{FEC}{forward error correction}
\acrodef{FFT}{fast Fourier transform}
\acrodef{FF}{fill factor}
\acrodef{FMS}{fixed mode selection}
\acrodef{FOV}{field of view}
\acrodefplural{FOV}[FOVs]{fields of view}
\acrodef{FSO}{free space optical}
\acrodef{GaAs}{gallium arsenide}
\acrodef{Gbps}{gigabits per second}
\acrodef{GEO}{geostationary earth orbit}
\acrodef{GoB}{grid-of-beam}
\acrodef{GPS}{global positioning system}
\acrodef{HAP}{high-altitude platform}
\acrodef{HARQ}{hybrid automatic repeat request}
\acrodef{HF}{high frequency}
\acrodef{HPBD}{half-power beam divergence}
\acrodef{HPBW}{half-power beamwidth}
\acrodef{IBI}{inter-beam interference}
\acrodef{IEC}{International electrotechnical commission}
\acrodef{IF}{intermediate-frequency}
\acrodef{IFFT}{inverse fast Fourier transform}
\acrodef{ISL}{inter-satellite link}
\acrodef{ITS}{intelligent transportation system}
\acrodef{JT}{joint transmission}
\acrodef{IM{-}DD}{intensity modulation and direct detection}
\acrodef{IOP}{institute of photonics}
\acrodef{IoT}{internet of things}
\acrodef{IoUT}{internet of underwater things}
\acrodef{IR}{infrared}
\acrodef{ISI}{inter-spot interference}
\acrodef{ICI}{inter-cluster interference}
\acrodef{IM{/}DD}{intensity modulation and direct detection}
\acrodef{JT}{joint transmission}
\acrodef{LAP}{low-altitude platform}
\acrodef{laser}{light amplification by stimulated emission of radiation}
\acrodef{LC}{liquid crystal}
\acrodef{LD}{laser diode}
\acrodef{LED}{light-emitting diode}
\acrodef{LEO}{low earth orbit}
\acrodef{LiFi}{light fidelity}
\acrodef{LO}{local oscillator}
\acrodef{LOS}{line-of-sight}
\acrodef{LRDC}{\ac{LiFi} research and development centre}
\acrodef{LTE}{long-term evolution}
\acrodef{Mbps}{megabits per second}
\acrodef{MBS}{multicast broadcasting services}
\acrodef{MHP}{most hazardous position}
\acrodef{MIMO}{multiple-input multiple-output}
\acrodef{MJ}{multi-junction}
\acrodef{ML}{machine learning}
\acrodef{MLE}{maximum likelihood estimator}
\acrodef{MPE}{maximum permissible exposure}
\acrodef{MPP}{maximum power point}
\acrodef{MPPC}{multi-pixel photon counter}
\acrodef{MRC}{maximum ratio combining}
\acrodef{MRR}{modulating retro-reflectors}
\acrodef{MUI}{multi-user interference}
\acrodef{NoN}{network of networks}
\acrodef{NIR}{near infrared}
\acrodef{NLOS}{non-line-of-sight}
\acrodef{NOMA}{non-orthogonal multiple access}
\acrodef{NR}{new radio}
\acrodef{NTN}{non-terrestrial network}
\acrodef{OAM}{orbital angular momentum}
\acrodef{OFDM}{orthogonal frequency division multiplexing}
\acrodef{OFDMA}{orthogonal frequency division multiple access}
\acrodef{OLED}{organic \ac{LED}}
\acrodef{OPA}{optical phased array}
\acrodef{OPD}{organic photodetector}
\acrodef{OPV}{organic \ac{PV}}
\acrodef{ORIS}{optical reconfigurable intelligent surface}
\acrodef{OW}{optical wireless}
\acrodef{OWC}{optical wireless communication}
\acrodef{OI}{optical interconnect}
\acrodef{PA}{power amplifier}
\acrodef{PAM}{pulse amplitude modulation}
\acrodef{PAT}{pointing, acquisition and tracking}
\acrodef{PCE}{power conversion efficiency}
\acrodef{PDCCH}{physical downlink control channel}
\acrodef{PIN}{positive-intrinsic-negative}
\acrodef{PMT}{photomultiplier tube}
\acrodef{PSD}{power spectral density}
\acrodef{PD}{photodiode}
\acrodef{PPC}{photonic power converter}
\acrodef{PV}{photovoltaic}
\acrodef{QAM}{quadrature amplitude modulation}
\acrodef{QKD}{quantum key distribution}
\acrodef{QoS}{quality-of-service}
\acrodef{RedCap}{reduced capability}
\acrodef{RF}{radio frequency}
\acrodef{RGB}{red-green-blue}
\acrodef{RIS}{reconfigurable intelligent surfaces}
\acrodef{RIN}{relative intensity noise}
\acrodef{RIS}{reconfigurable intelligent surface}
\acrodef{RNN}{recurrent neural network}
\acrodef{RSS}{received signal strength}
\acrodef{Rx}{receiver}
\acrodef{SBC}{static beam clustering}
\acrodef{SDMA}{space division multiple access}
\acrodef{SIC}{successive interference cancellation}
\acrodef{SINR}{signal-to-interference-plus-noise ratio}
\acrodef{SLIPT}{simultaneous lightwave information and power transfer}
\acrodef{SLM}{spatial light modulator}
\acrodef{SNR}{signal-to-noise ratio}
\acrodef{SPAD}{single photon avalanche detector}
\acrodef{STAR}{simultaneous transmit and reflect}
\acrodef{SVD}{singular value decomposition}
\acrodef{TDoA}{time difference of arrival}
\acrodef{TEM}{transverse electromagnetic}
\acrodef{THz}{tera-Hertz}
\acrodef{TIA}{transimpedance amplifier }
\acrodef{ToA}{time of arrival}
\acrodef{ToR}{top-of-rack}
\acrodef{Tx}{transmitter}
\acrodef{TV}{television}
\acrodef{UAC}{underwater acoustic communication}
\acrodef{UE}{user equipment}
\acrodef{UEC}{underwater electromagnetic communication}
\acrodef{UHD}{ultra-high-definition}
\acrodef{URLLC}{ultra-reliable low-latency communication}
\acrodef{UV}{ultraviolet}
\acrodef{UWOC}{underwater wireless optical communication}
\acrodef{UWSN}{underwater wireless sensor network}
\acrodef{V2X}{vehicle-to-everything}
\acrodef{V2V}{vehicle-to-vehicle}
\acrodef{V2I}{vehicle-to-infrastructure}
\acrodef{V2P}{vehicle-to-pedestrian}
\acrodef{VRU}{vulnerable road user}
\acrodef{V2N}{vehicle-to-network}
\acrodef{ITS}{intelligent transportation system}
\acrodef{C2X}{car-to-everything}
\acrodef{DSRC}{dedicated short-range communication}
\acrodef{OWC}{optical wireless communication}
\acrodef{NOMA}{non-orthogonal multiple access}
\acrodef{6G}{sixth-generation}
\acrodef{C-V2X}{cellular vehicle-to-everything}
\acrodef{LTE}{long-term evolution}
\acrodef{RF}{radio frequency}
\acrodef{C-V2X}{cellular vehicle-to-everything}
\acrodef{URLLC}{ultra-reliable low-latency communication}
\acrodef{AI}{artificial intelligence}
\acrodef{NR}{new radio}
\acrodef{IRS}{intelligent reflective surface}
\acrodef{ML}{machine learning}
\acrodef{QR}{quantum computing}
\acrodef{VCSEL}{vertical cavity surface emitting laser}
\acrodef{VLC}{visible light communication}
\acrodef{VR}{virtual reality}
\acrodef{WAVE}{wireless access in vehicular environments}
\acrodef{WDM}{wavelength division multiplexing}
\acrodef{WiFi}{wireless fidelity}

\begin{document}
\title{Optical Wireless Communications: Enabling the Next Generation Network of Networks}

\author{%
Aravindh~Krishnamoorthy\textsuperscript{\orcidlink{0000-0001-7186-121X}}\IEEEauthorrefmark{1},~\IEEEmembership{Member,~IEEE,} %
Hossein~Safi\textsuperscript{\orcidlink{0000-0002-5427-7780}}\IEEEauthorrefmark{1}, ~\IEEEmembership{Member,~IEEE,} %
Othman~Younus\textsuperscript{\orcidlink{0000-0003-2015-1175}}\IEEEauthorrefmark{1},~\IEEEmembership{Member,~IEEE,} %
Hossein~Kazemi\textsuperscript{\orcidlink{0000-0002-5051-0565}}\IEEEauthorrefmark{1},~\IEEEmembership{Member,~IEEE,} %
Isaac~N.~O.~Osahon\textsuperscript{\orcidlink{0000-0001-5685-6068}}\IEEEauthorrefmark{1},~\IEEEmembership{Member,~IEEE,} %
Mingqing~Liu\textsuperscript{\orcidlink{0000-0002-5658-7811}}\IEEEauthorrefmark{1},~\IEEEmembership{Member,~IEEE,} %
Yi~Liu\IEEEauthorrefmark{1}, %
Sina~Babadi\IEEEauthorrefmark{1}, %
Rizwana~Ahmad\IEEEauthorrefmark{1}, %
Asim~Ihsan\IEEEauthorrefmark{1}, %
Behnaz~Majlesein\textsuperscript{\orcidlink{0000-0001-5009-3987}}\IEEEauthorrefmark{1}, %
Yifan~Huang\IEEEauthorrefmark{1}, %
Johannes~Herrnsdorf\IEEEauthorrefmark{2}, %
Sujan~Rajbhandari\textsuperscript{\orcidlink{0000-0001-8742-118X}}\IEEEauthorrefmark{2},~\IEEEmembership{Senior Member,~IEEE,} %
Jonathan~J.D.~McKendry\textsuperscript{\orcidlink{0000-0002-6379-3955}}\IEEEauthorrefmark{2},~\IEEEmembership{Member,~IEEE,} %
Iman~Tavakkolnia\IEEEauthorrefmark{1},~\IEEEmembership{Member,~IEEE,} %
Humeyra~Caglayan\textsuperscript{\orcidlink{0000-0002-0656-614X}}\IEEEauthorrefmark{3}, %
Henning~Helmers\textsuperscript{\orcidlink{0000-0003-1660-7651}}\IEEEauthorrefmark{4},~\IEEEmembership{Senior Member,~IEEE,} %
Graham~Turnbull\textsuperscript{\orcidlink{0000-0002-2132-7091}}\IEEEauthorrefmark{5},~\IEEEmembership{Senior Member,~IEEE,} %
Ifor~D.~W.~Samuel\textsuperscript{\orcidlink{0000-0001-7821-7208}}\IEEEauthorrefmark{5}, %
Martin~D.~Dawson\textsuperscript{\orcidlink{0000-0002-6639-2989}}\IEEEauthorrefmark{2},~\IEEEmembership{Fellow,~IEEE,} %
Robert~Schober\IEEEauthorrefmark{6},~\IEEEmembership{Fellow,~IEEE,} %
Harald~Haas\IEEEauthorrefmark{1},~\IEEEmembership{Fellow,~IEEE} \\
{%
\bigskip
\small
*~LiFi Research and Development Centre, Electrical Engineering Division, University of Cambridge, Cambridge, UK,\\
\dag~Institute of Photonics, Department of Physics, University of Strathclyde, UK,\\
\ddag~Faculty of Engineering and Natural Science, Photonics, Tampere University, Finland,\\
\textsection~Fraunhofer Institute for Solar Energy Systems ISE, Freiburg im Breisgau, Germany,\\
$\scriptstyle\mathparagraph$~Organic Semiconductor Centre, School of Physics and Astronomy, University of St. Andrews, UK,\\
\textbardbl~Institute for Digital Communications, Friedrich-Alexander-Universität Erlangen-Nürnberg, Germany\\
}~%
\thanks{This work was supported by the Future Telecoms Research Hub, Platform for Driving Ultimate Connectivity (TITAN), sponsored by the Department of Science Innovation and Technology (DSIT) and the Engineering and Physical Sciences Research Council (EPSRC) under Grant EP/X04047X/1 and Grant EP/Y037243/1.}%
\thanks{Humeyra~Caglayan acknowledges the support of CHIST-ERA grant CHIST-ERA-21-NOEMS-002, by the Research Council of Finland 357746.}%
\thanks{Graham Turnbull acknowledges the support of UKRI grant reference 10120583.}
\thanks{The work of Robert~Schober was partly supported by the Deutsche Forschungsgemeinschaft (DFG, German Research Foundation) under grant SCHO 831/17-1.}%
\thanks{Harald~Haas and Rizwana~Ahmad acknowledge support from the EPSRC under CHIST-ERA grant EP/X034542/2 (Meta-LiFi), and Harald~Haas and Hossein~Kazemi acknowledge support by the Engineering and Physical Sciences Research Council (EPSRC) under grant EP/S016570/1 `Terabit Bidirectional Multi-User Optical Wireless System (TOWS) for 6G LiFi'.}}

\markboth{Journal of \LaTeX\ Class Files,~Vol.~14, No.~8, August~2015}%
{Shell \MakeLowercase{\textit{et al.}}: Bare Demo of IEEEtran.cls for IEEE Journals}

\IEEEspecialpapernotice{(Invited Paper)}

\maketitle

\begin{abstract}
Optical wireless communication (OWC) is a promising technology anticipated to play a key role in the next-generation network of networks, especially as a complementary technology to traditional radio frequency communications, for enhancing networking capabilities beyond conventional terrestrial networks. OWC is already a mature technology with diverse usage scenarios, and can enable integrated applications via wireless access and backhaul networks, dynamic drone and satellite networks, underwater networks, inter- and intra-system interconnecting networks, and vehicular communication networks. Furthermore, novel and emerging technological opportunities such as photovoltaic cells, orbital angular momentum-based modulation, optical reconfigurable intelligent surfaces, organic light-emitting and photo diodes, and recent advances in ultraviolet communications can help enhance future OWC capabilities even further. Moreover, OWC networks can also support value-added services such as enhanced positioning and gesture recognition. Hence, OWC provides unique functionalities that can play a crucial role in building convergent and resilient future network of networks alongside radio frequency and optical fiber technologies.
\end{abstract}

\IEEEpeerreviewmaketitle

\section{Introduction} \label{sec:intro}
\IEEEPARstart{T}{he} first demonstrations of \ac{laser} and \acp{LED} in the $1960$s laid the foundations for what would be the rich and promising field of \ac{OWC}. In the $1970$s, these devices led to the development of the first \ac{OWC} links for ground-to-satellite communications at NASA, and \ac{IR}-based wireless computer networks at IBM research labs. However, despite this early success, most of the explosive growth in \ac{OWC} research and development has only occurred in the last two decades, driven by the exponential increase in mobile data use, unprecedented connectivity requirements of fully integrated \ac{NoN}, such as the \ac{4G} networks, and availability of a wide variety of off-the-shelf optical components.

Today, \ac{OWC} is an established technology with applications ranging from Starlink's inter-satellite interconnects to IEEE~802.11bb-standardized indoor wireless networks, and a promising future growth potential owing to its unique advantages, which include: a vast available unlicensed spectrum, very high physical layer security, resilience to fading in \ac{IM{/}DD} systems, and a mature optical device ecosystem, thanks to the developments in optical fiber networks.

\begin{figure}
	\centering
	\includegraphics[width=0.95\linewidth]{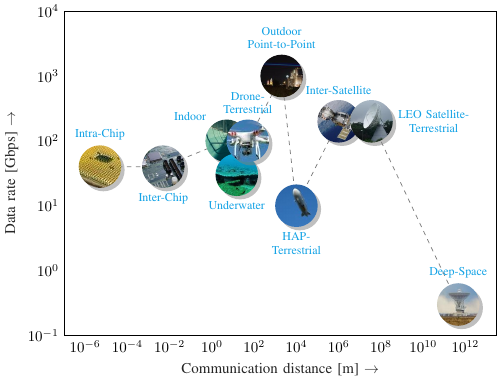}
	\def\mycaptiontext{The scale of \ac{OWC}-based links ranges from intra-chip interconnects, replacing complex bus structures, to deep-space communications, and achieving data rates of up to Tbps}
	\captionof{figure}[\mycaptiontext.]{\mycaptiontext\protect\footnotemark.}
	\label{fig:scale}
\end{figure}

On the other hand, the recent surge in mobile data consumption, the \emph{digital divide} caused by uneven access to digital technologies, and the need for networks with low energy consumption are factors that are shaping the evolution of future connectivity. Furthermore, a shift towards new standards such as \ac{6G} and beyond is accelerating the unification of heterogeneous networks, such as mobile, satellite, aerial, underwater, and \ac{IoT} networks, into a single intelligent \ac{NoN}, demanding unprecedented inter-networking capabilities. Fortunately, \ac{OWC} is well positioned to meet these requirements due to its wireless nature and tremendous scalability, which ranges from intra-chip interconnects to deep-space communications, see Fig. \ref{fig:scale} on the next page, as well as its ability to complement technologies such as \ac{RF} and optical fiber.

\footnotetext{Sources -- Intra-Chip: J. Xue et al., ACM SIGARCH, 2010; Inter-Chip: R. Baets et al., Opt. Mat., 2001; Indoor: \cite{Cheng2024100Gbps}; Underwater: WS. Tsai et al., Sci. Rep., 2019; Drone-Terrestrial: S. M. Walsh et al., Sci. Rep., 2022; Outdoor Point-to-Point: M. A. Fernandes et al., J. Lightwave Tech., 2024; HAP-Terrestrial: F. Fidler et al., IEEE JSTQE, 2010; Inter-Satellite: SpaceX, 2024; LEO Satellite-Terrestrial: NASA TBIRD, 2023; Deep-Space: NASA Psyche, 2024.}

However, implementing these \ac{NoN} \ac{OWC} links is not without challenges. It is well known that \ac{OWC} is best suited for \ac{LOS} conditions. Furthermore, mobility causes random misalignment and blockages, which can lead to connection losses (outages). Moreover, in outdoor deployments, atmospheric effects, such as turbulence, cause severe signal degradation. In this paper, several design strategies for overcoming these challenges are discussed. Novel and emerging solutions are presented. Furthermore, given the similarities in the application areas, a brief comparison of \ac{OWC} and \ac{THz} communication systems is provided. Interestingly, the solutions developed for \ac{OWC}, such as the \acp{GoB}, can also benefit \ac{THz} and other high-frequency \ac{RF} systems, which face similar challenges.

The remainder of this paper is organized as follows. In Subsection \ref{sec:comparisonTHz}, a brief comparison of \ac{OWC} and \ac{THz} communication systems is provided. Section~\ref{sec:networking} introduces key \ac{OWC} technologies that enable the \ac{NoN}, encompassing underwater, terrestrial, space, drone, and satellite networks. These technologies also augment existing \ac{RF} links for applications such as indoor access, wireless backhaul, and data centers. Additionally, intra- and inter-system \ac{OWC} links are discussed.

Next, in Section~\ref{sec:emerging}, emerging \ac{OWC} technologies that will enhance next generation \ac{OWC}-based \ac{NoN} across diverse domains, such as \acp{PV} for \ac{SLIPT}, \ac{OAM}-based \ac{OWC} for high-data-rate, reliable, and secure communication, \acp{ORIS} for controlling the propagation medium, \acp{OLED} and \acp{OPD} for developing low-cost and flexible devices, and novel \ac{UV} devices for outdoor \ac{OWC} communication in the solar-blind region, are introduced.

In addition to communication, light can also be exploited to enhance sensing technologies such as positioning and gesture recognition. In Section~\ref{sec:6gtech}, these \emph{value-added} technologies, which capitalize on \ac{OWC} integrated networks, are presented.

Furthermore, where applicable, potential research and development opportunities are described in an \emph{Opportunities} subsection, and key challenges and future research areas are outlined in a \emph{Challenges and Future Research Directions} subsection.

Lastly, Section~\ref{sec:conclusion} offers an outlook and concluding remarks.

\subsection{Comparison of \acs{OWC} and \acs{THz}} \label{sec:comparisonTHz}

Both \ac{THz} communications and \ac{OWC} are anticipated to have the potential to revolutionize next generation wireless communication systems. Therefore, comparing these technologies to identify their optimal application areas is of high interest. Several existing studies have provided qualitative comparisons between \ac{THz} and \ac{OWC} technologies, examining factors such as mobility support, blockage sensitivity, licensing requirements, and applicable IEEE standards, to highlight their respective advantages and challenges, see, e.g., \cite{THZ_compare}. While these high-level comparisons offer valuable insights, there still remains a need for theoretical modeling and quantitative analyses of the two technologies, which are presented below \cite{Liu2024}.

Among various \ac{OWC} technologies, \ac{VCSEL}-based \ac{OWC} is best positioned to enable \ac{LiFi} $2.0$ \cite{Soltani2023}, significantly contributing to key \ac{6G} targets such as high data density, ultra-high data rates, and low latency. On the other hand, electronic-\ac{THz} technology has been extensively researched and validated in semiconductor devices, showing miniaturization, cost-effectiveness, and good performance with easy integration into existing communication infrastructure compared to photonics- and plasmonic-based \ac{THz} technologies \cite{THZ_350GHz}. Hence, a quantitative comparison is presented in the following between 350 GHz electronic-\ac{THz} and 940 nm \ac{VCSEL}-based \ac{OWC} systems for indoor application scenarios in terms of their \ac{CF}, which is the maximum ratio of data rate to the power consumed. 

\subsubsection{\acs{OWC} and \acs{THz} Models}

In the \ac{VCSEL}-based \ac{OWC} system, the transmit signal is biased with \ac{DC} power using a bias tee at the transmitter, which then jointly drive the \ac{VCSEL} to emit information-carrying beams. After propagating through free space, the beam power is captured by a \ac{PD} at the receiver and amplified by a \ac{TIA}. Our system model \cite{Liu2024} encompasses: (i) non-linear electro-optic conversion of the \ac{VCSEL} at the transmitter, (ii) Gaussian beam propagation-based channel modeling with misalignment, and (iii) noise modeling at the receiver. Furthermore, additional challenges such as bandwidth limitations, non-negativity constraints, and compliance with eye-safety regulations for transmission power are taken into account in our analysis.

On the other hand, in line with literature, the electronic-\ac{THz} communication systems are modeled as cascaded system comprising a \ac{PA}, \ac{BPF}, mixer, and \ac{LO}. In our model, at the transmitter, the \ac{IF} signal is amplified by the PA, filtered through the \ac{BPF}, mixed with the LO signal, and then converted into a \ac{HF} signal for emission via an antenna. The receiver performs the reverse process, converting the \ac{HF} signal back to \ac{IF} for demodulation. Furthermore, we model the the wave propagation in indoor environments at $350$ GHz via a multi-ray channel model, and incorporate the effects of phase noise. In addition, for computing the overall power consumption, we consider the power consumed by the individual components such as the \ac{LO}.

\subsubsection{Comparison}

High-frequency propagation incurs significant path loss, which necessitates the use of high-gain antennas for both \ac{VCSEL}-based \ac{OWC} and \ac{THz} communication systems. Specifically, optical lenses are employed for \ac{VCSEL}-based \ac{OWC}, while high-directivity antennas are used in \ac{THz} systems. This shared need for high-gain antennas introduces a similar tradeoff in both systems: improving channel gain often limits mobility, as directional antennas require stable alignment for optimal performance. By maintaining the same \ac{HPBW} for the \ac{THz} antenna and \ac{HPBD} for the laser post-lens, the energy efficiency comparison can then account for mobility constraints. 

\begin{figure}
	\centering
	\includegraphics[width=0.9\linewidth,keepaspectratio]{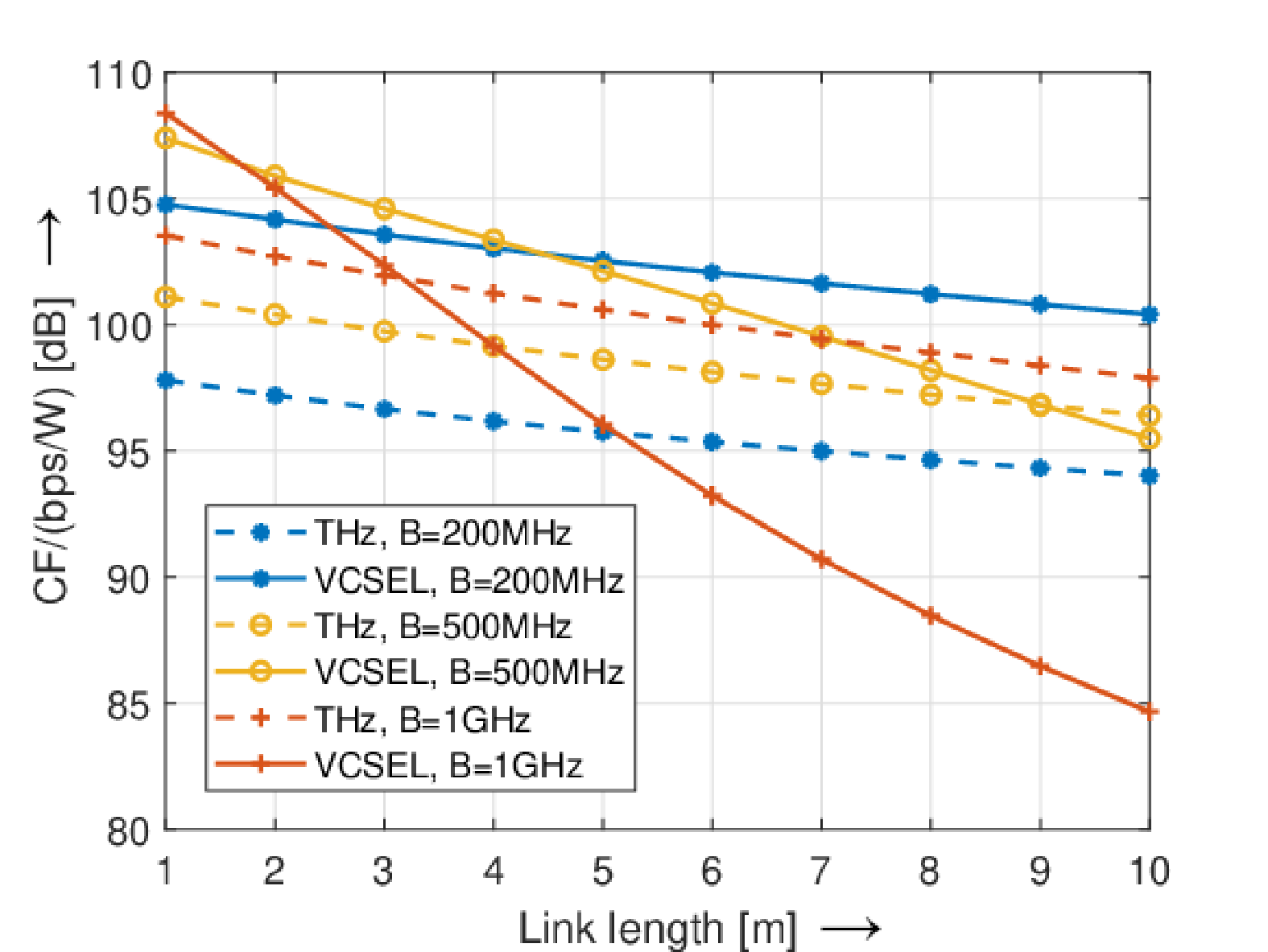}
	\caption{\ac{CF} comparison of \ac{THz} and \ac{VCSEL}-based \ac{OWC} systems under varying link length with different system bandwidth.}
	\label{fig:cfcompaligned}
\end{figure}

In Fig. \ref{fig:cfcompaligned}, the impact of system bandwidth and transmission distance are shown. Here, we configure both \ac{HPBW} and \ac{HPBD} to $4^\circ$ for a fair comparison. Compared to \ac{THz} communications, we find that the \ac{VCSEL}-based system is more sensitive to link length, especially for higher bandwidths. That is, for a bandwidth of $B=200$ MHz, the \ac{VCSEL}-based system demonstrates significant \ac{CF} advantages. However, at a higher bandwidth of $B=1$ GHz, \ac{VCSEL}-based \ac{OWC} outperforms the \ac{THz} system only within a transmission distance of $3$ m. Besides, during our simulations, we observed that beam focusing is crucial for achieving the optimal performance of \ac{VCSEL}-based systems.

\section{Key \acs{OWC} Network Technologies}
\label{sec:networking}

In this section, we present key \ac{OWC} network technologies that can complement \ac{RF} and fiber optics and enable next-generation \acp{NoN}.

\subsection{Wireless Access Networks}

Today's societies are becoming increasingly data centric with millions of devices and sensors planned to be integrated into smart environments such as cities, homes, workplaces, and production facilities, driving massive data volumes at unprecedented speeds \cite{Giordani2020toward6G} and necessitating a paradigm shift in wireless access network design.

To this end, there have been significant standardization efforts in the area of optical wireless networking, notably the IEEE802.11bb and ITU-T G.hn standards. The IEEE802.11bb standard operates in the wavelength range 800~nm to 1000~nm, whereas the ITU-T G.hn standard is a a more general home networking standard and does not specify a spectrum range. 

\begin{figure*}
     \centering
     \begin{minipage}[t]{0.9\textwidth}
     \centering
     \subfloat[\label{Fig:GoB_Coverage} \acs{AP} architecture and \acs{GoB} coverage]{\includegraphics[width=0.4\textwidth, keepaspectratio=true]{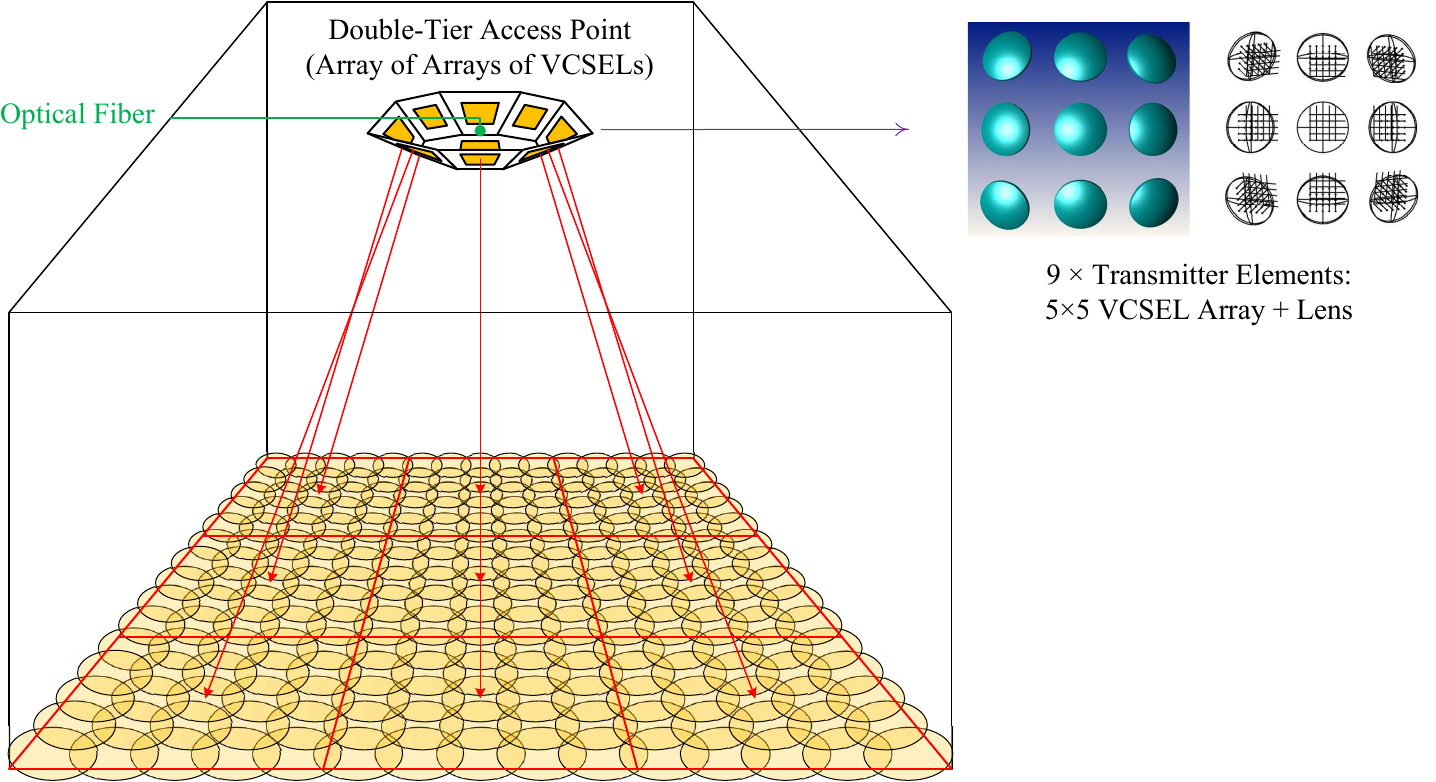}}
     \hfill
     \centering
     \subfloat[\label{Fig:SBC_SINR} \acs{SINR} for \acs{SBC} with CoMB-JT]{\includegraphics[width=0.3\textwidth, keepaspectratio=true]{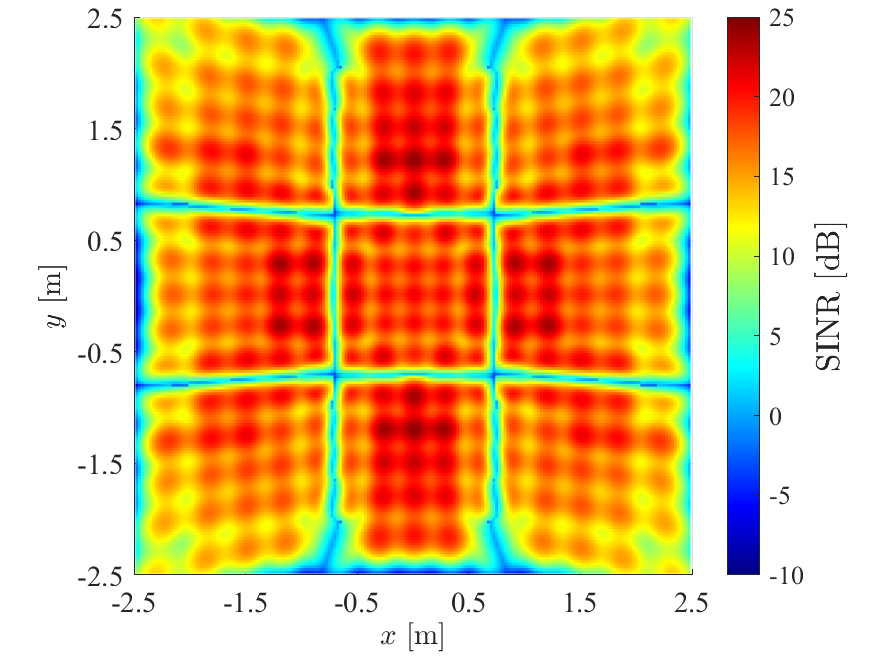}}
     \hfill
     \centering
     \subfloat[\label{Fig:DBC_SNR} \acs{SNR} for \acs{DBC} with CoMB-JT]{\includegraphics[width=0.3\textwidth, keepaspectratio=true]{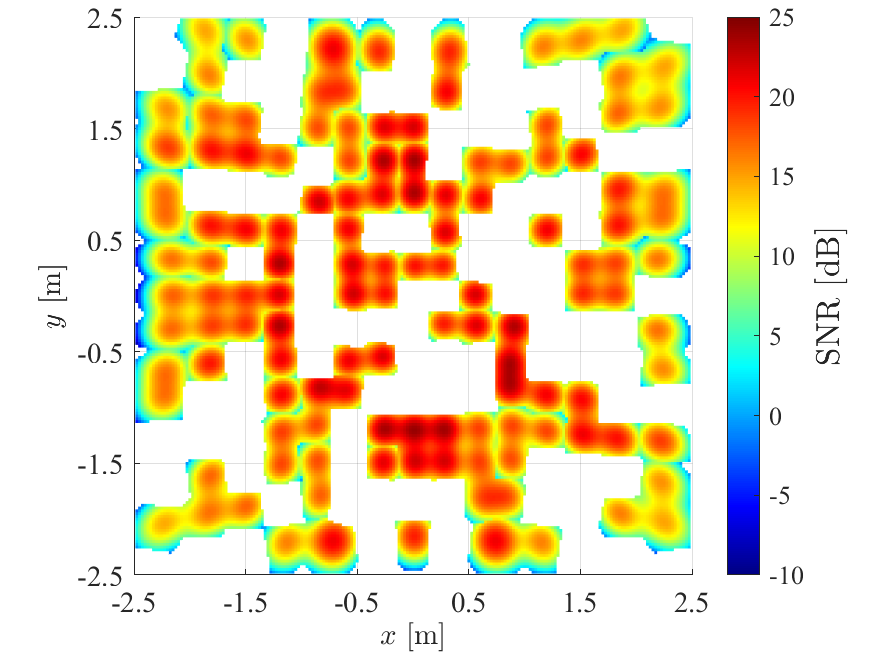}}
    \end{minipage}
    \caption{Indoor \acf{GoB} optical wireless access network using double-tier \acs{AP} design based on an array of arrays of \acs{VCSEL}. The \acf{SBC} scenario shown in (b) uses 9 fixed clusters, each of size 25. The \acf{DBC} scenario shown in (c) is based on $100$ \acp{UE} randomly distributed in the network; the beams are activated to only cover the regions where the \acp{UE} are located.}
	\label{Fig:TOWS_Architecture}
\end{figure*}

\subsubsection{Opportunities}

\Ac{OWC} technology based on \acp{laser}, leveraging narrow-beam \ac{IR} lasers, has demonstrated significant potential for enabling ultra-high-speed data transmission in indoor multi-user wireless networks \cite{Koonen2020ultra}. A novel optical wireless \ac{AP} architecture using \ac{VCSEL} arrays has been proposed for \ac{6G} \ac{LiFi} networks to achieve data rates greater than $10$~Gb/s per \ac{UE} with seamless indoor coverage \cite{Sarbazi2020TOWS_AP,Kazemi2024TOWS_GoB}. This architecture, also referred to as double-tier \ac{AP}, is based on an array of arrays of \acp{VCSEL} to provide \ac{GoB} coverage, as depicted Fig.~\ref{Fig:TOWS_Architecture}.

According to this \ac{GoB} approach, beams are massively deployed and are allowed to be partially overlapping, leading to \ac{IBI}. For \ac{IBI} management in the proposed \ac{GoB} \ac{LiFi} network, novel schemes based on \ac{AI} and \ac{ML} can be utilized. On the other hand, schemes inspired by \ac{CoMP} for \ac{RF}, such as \ac{CoMB}-\ac{JT} can also be deployed. Below, we briefly describe \ac{CoMB}-\ac{JT}, through which, we reveal the challenges involved in \ac{IBI} management.

In \ac{CoMB}-\ac{JT} with \ac{SBC}, a fixed number of clusters using a predefined layout for clustered beams are employed to enhance the received \ac{SINR} of the \acp{UE}. %
However, the performance of the low-complexity \ac{SBC} approach is limited by \ac{ICI} owing to the fixed number and layout of the clusters. On the other hand, in \ac{CoMB}-\ac{JT} with \ac{DBC}, the clustering layout is adaptively changed in response to network variations, e.g., based on the position of the \acp{UE}, see Section \ref{sec:indoorpositioning} for details on precise positioning of \acp{UE} based on \ac{OWC}. This approach ensures perfect avoidance of \ac{ICI} by dynamically forming clusters, so that \acp{UE} are always positioned within \ac{ICI}-free regions, while narrow interference zones are confined to user-free areas of the room. This is achievable due to the spatial confinement of light beams, which enhances both the performance per \ac{UE} and the overall network capacity.

\subsubsection{Challenges and Future Research Directions}

The main challenges focus on overcoming the issues related to the \ac{LOS} requirement for high-speed communication links. This is particularly important for mobile and multi-user applications. 

\textbf{\ac{LOS}: Mobile wireless communications rely on the fact that transmitters and receivers do not need to point directly at each other. In contrast, optical wireless links are highly directional. This presents a challenge that can, however, be addressed through the appropriate design of optical transmitters and receivers. For example, this can involve adopting principles inspired by the insect facet eye, using angular diversity in optical receivers and transmitters.}

{\textbf{Optical gain-bandwidth trade-off due to conservation of \'etendue:}\label{sec:etendue}}
In order to achieve high electrical bandwidth, small-area photodetectors are required. In order to ensure sufficient photons are captured to achieve a given \ac{SNR}, the system relies on either imaging optics, e.g., lenses, or non-imaging optics, e.g., concentrators, to improve the photon collection performance. However, as a consequence of the law of conservation of \'etendue, which is derived from the law of conservation of energy in optical systems, higher optical gains lead to lower \acp{FOV}, which poses a challenge in the context of non-directional networking.

\textbf{Optical-to-electrical conversion efficiency:}
The process of converting electrical signals into optical signals at the transmitter, typically using devices such as \acp{laser} or \acp{LED}, and subsequently converting the optical signals back into electrical signals at the receiver using photodetectors such as \acp{APD} or \ac{PIN} diodes, imposes a critical bottleneck. This bottleneck arises due to the limited bandwidth conversion efficiency of these devices. Consequently, the system performance is constrained, preventing full utilization of the vast optical spectrum available. To address this limitation, it is imperative to enhance the electrical bandwidth of these devices. Achieving higher electrical bandwidth not only improves conversion efficiency but also unlocks the potential for leveraging the enormous data-carrying capacity of optical communication systems.

\subsection{Wireless Backhaul Networks}

Next, we consider optical wireless backhaul, which have an immense potential to transform backhauling in point-to-point networks as well as in integrated communication systems such as \ac{6G} and beyond systems.

\subsubsection{Opportunities}

Although widely utilized, conventional \emph{wired} backhaul networks, based on Ethernet and fiber-optic links, incur significant infrastructure costs and lack deployment flexibility. On the other hand, conventional \ac{RF}-based \emph{wireless} backhaul networks suffer from a shortage of available bandwidth in the contentious \ac{RF} spectrum. These shortcomings have driven the development of optical wireless backhaul networks, which enable high-data-rate, low-latency, secure communication by exploiting the large license-free bandwidth available in visible or infra-red regions and are especially relevant for short-range backhauling.

Several optical backhaul designs have been proposed in the literature, see \cite{Tezergil2022WirelessBackhaul} and references therein. On the experimental side, an indoor $100$ Gbps \ac{WDM}-based system is demonstrated in \cite{Cheng2024100Gbps}. As well, outdoor links of up to $4.8$ Gbps for distances of up to $500$ meters with diffuse Lambertian lasers and low-cost \acp{LED} are demonstrated in \cite{Cheng2024100Gbps} and \cite{Schulz2016RobustOptical}, respectively. The demonstrated designs are well suited for indoor and outdoor optical wireless backhauling.

\begin{figure*}
	\centering
	\begin{minipage}[t]{0.48\textwidth}
		\includegraphics[clip, trim=0cm 0cm 11.75cm 0cm, width=0.9\linewidth, keepaspectratio=true]{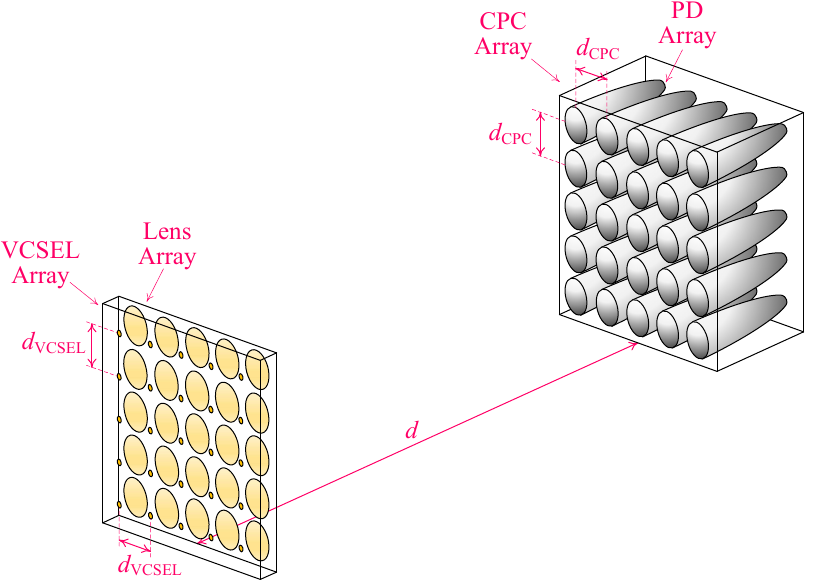}
		\caption{Terabit MIMO OWC link design using a VCSEL array and a multi-element non-imaging receiver based on CPCs. Here, $d_\mathrm{VCSEL}$ and $d_\mathrm{CPC}$ refer to the inter-\ac{VCSEL} and inter-\ac{PD} distances at the transmitter and receiver, respectively, and $d$ denotes the communication distance.}
		\label{fig:mimo-owc}
	\end{minipage}%
	\hspace{0.04\textwidth}%
	\begin{minipage}[t]{0.48\textwidth}
		\includegraphics[width=0.9\linewidth]{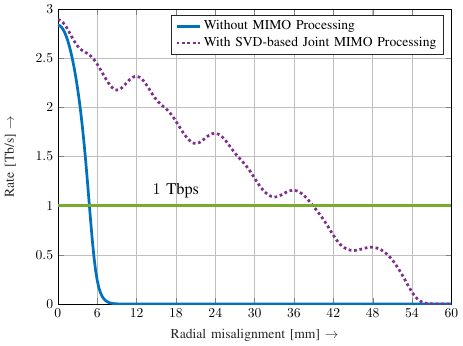}
		\caption{Throughput vs. radial misalignment distance for a \ac{MIMO} optical wireless system with $N_\mathrm{t} = 25$ $(5\times 5)$ transmit \ac{VCSEL} array with $6$ mm inter-element spacing and a receiver \ac{PD} array with $N_\mathrm{r} = 81$  \acp{PD}, with and without \ac{SVD}-based joint signal processing. Figure from \cite{Kazemi2022ATbps}.}
		\label{fig:wb-1}
	\end{minipage}
\end{figure*}

\subsubsection{Challenges and Future Research Directions} 

Despite their immense potential, optical wireless backhaul and data-center networks inherit several design challenges of \acp{OWC}.

\textbf{Atmospheric effects:} Specifically, for wireless backhaul, careful system design for transmitter--receiver alignment and overcoming impairments such as turbulence, fog, and rain are necessary. To this end, the use of hybrid optical--\ac{RF} backhaul systems, which address some of the above concerns, seem promising.

\textbf{Misalignment:} A key practical challenge in designing narrow laser-beam-based optical wireless links lies in their susceptibility to misalignment errors. To address link misalignment, the use of multi-element non-imaging receiver design incorporating \acp{CPC} \cite{Sarbazi2024Receiver} is interesting owing to its large field of view and high concentration gain.

\textbf{\Ac{MIMO}:} Most of the aforementioned systems employ a single transmitter and receiver. Leveraging the \ac{MIMO} technology by utilizing a \ac{GoB} is promising for enhancing the data rate even further \cite{Kazemi2022ATbps}. 
In addition to the increased data rate, \ac{MIMO} systems can take advantage of \ac{SVD}-based joint \ac{MIMO} signal processing to improve the robustness to optical misalignment between transmitter and receiver. For example, the \ac{VCSEL}-array-based terabit MIMO OWC system in Fig. \ref{fig:mimo-owc} can tolerate a radial misalignment of up to six times the inter-element spacing at the transmitter while still satisfying a minimum data rate requirement of $1$ Tbps, see Fig. \ref{fig:wb-1}.

\subsection{Drone and Satellite Networks} 

In the vision for 6G and beyond, aerial and space platforms, i.e., \acp{LAP}, \acp{HAP}, and satellites, are expected to become essential components of the global \ac{NoN}, especially for enabling connectivity to under-served and remote areas. However, to fully realize their capabilities, communication links that provide high throughput, security, and low latency -- all key features of \ac{OWC} -- are crucial. 

\subsubsection{Opportunities}

\ac{OWC}-based links are particularly well-suited for \acp{HAP} and \acp{LAP}, where traditional RF systems often face limitations, as they provide point-to-point, high-bandwidth links that can support, e.g., applications like autonomous navigation, disaster relief, temporary networks, last-mile connectivity, smart city infrastructure monitoring, and immersive augmented reality experiences. Especially for \acp{LAP}, unlike \ac{RF} links, \ac{OWC} links are virtually interfere-free owing to their \ac{LOS} nature, thereby enabling high-density operation. On the other hand, in satellite communications as well, \ac{OWC} is a promising alternative to RF, particularly in \ac{LEO} constellations, where \ac{RF} systems struggle to meet the demand for high data rates due to their limited capacity and radio interference. \ac{OWC} systems also require less power for their operation. As well, \ac{OWC} offers significant benefits for \acp{ISL} within satellite constellations.

\begin{figure*}
	\centering
	\colorbox{gray!20}{%
		\begin{minipage}{0.2\linewidth}
			\scriptsize
			\begin{center}
				\textbf{Challenges}
			\end{center}
			\textbf{Space-Based Links}
			\begin{itemize}
				\item Alignment under satellite 
				movement.
				\item Short communication intervals.
				\item Compact, low-power equipment.
				\item SWaP limits on CubeSats.
				\item Cooling for high-power optics.
			\end{itemize}
			\textbf{Aerial Links}
			\begin{itemize}
				\item Power allocation and alignment for multiple users.
				\item Seamless handover for moving drones.
				\item Lightweight efficient systems.
				\item Vibration sensitivity.
				\item SWaP limits on drones.
			\end{itemize}
			\textbf{Network Integration for \ac{6G}}
			\begin{itemize}
				\item Coordinating with \ac{6G} infrastructure.
				\item Reducing delays.
				\item Limited link budget.
			\end{itemize}
	\end{minipage}}%
	\hspace{0.02\linewidth}%
	\begin{minipage}{0.75\linewidth}
		\centering
		\includegraphics[clip, trim=8.3cm 0cm 0cm 0cm, width=\linewidth]{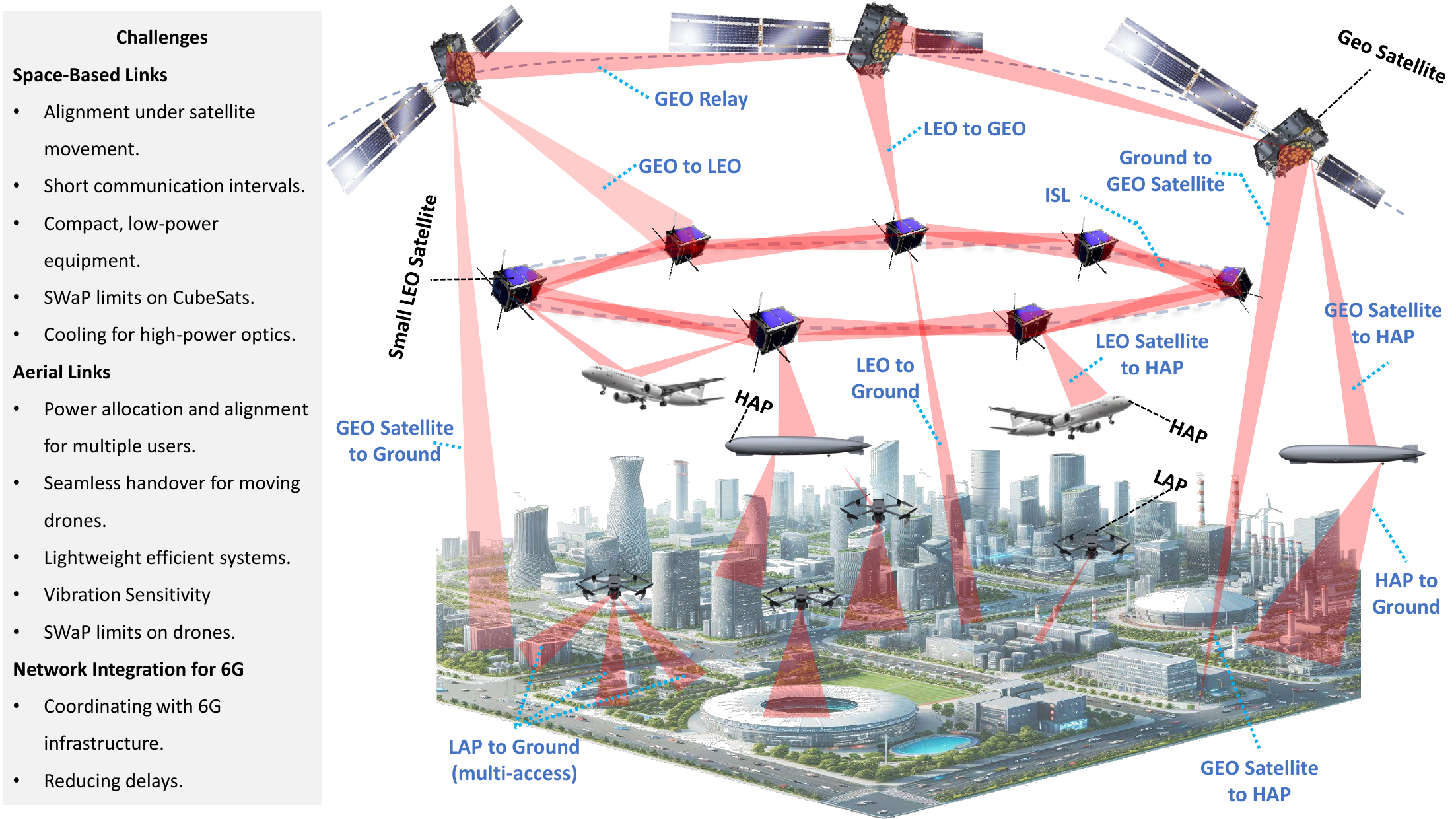}
	\end{minipage}
	\caption{Potential scenarios of aerial/space-based OWC links in 6G and beyond. SWaP: Size, weight, and power.}
	\label{fig:enter-label-aerail}
\end{figure*}

\subsubsection{Challenges and Future Research Directions}
As depicted in Fig. \ref{fig:enter-label-aerail}, there are numerous potential deployment scenarios for aerial- and space-based \ac{OWC} links in \ac{6G} and beyond. However, several significant challenges must be addressed in order to realize the full potential of such links, which are outlined in the following.

\textbf{Beam pointing and acquisition:} Maintaining precise beam pointing and acquisition becomes crucial with narrow beams under the high-speed, high-vibration conditions of \acp{ISL}. Furthermore, environmental factors such as atmospheric turbulence affect downlink transmission. Constant tracking and adjustment of \ac{AoA} and \ac{AoD} is often necessary to maintain stable links. However, size, weight, and power limitations make the design of such adaptive subsystems challenging.

\textbf{Miniaturized terminals:} Small \ac{LEO} satellites, like CubeSats, and drones face strict size, weight, and power constraints that limit the types of \ac{OWC} equipment they can carry, necessitating compact, low-power \ac{OWC} devices. For these applications, adaptations of technologies like \ac{VCSEL} arrays, which have been typically used for indoor communications, seem promising.

\textbf{Multi-user access points:} Supporting multiple simultaneous \ac{OWC} connections from a single drone is challenging due to the limited onboard processing power, battery life, alignment requirements, and the need for efficient power distribution among all users. Furthermore, ensuring stable links while the drone is in motion requires advanced multi-link management and dynamic beam steering. Recently, the use of \ac{RIS} and \ac{MRR} mounted on drones have been explored to overcome these challenges.

\textbf{Handover:} Fast-moving platforms pose difficulties in maintaining stable \ac{OWC} connections, often necessitating frequent handovers, especially in high-mobility environments, which can disrupt connectivity and increase latency. Therefore, efficient handover mechanisms are crucial for seamless \ac{OWC} communication in such scenarios.

\textbf{Energy efficiency for prolonged operation:} Drones and small satellites are energy-constrained which makes energy efficiency a priority. Maintaining high-throughput \ac{OWC} links while minimizing power consumption requires optimized systems that are both lightweight and less complex. Power-efficient beam acquisition and tracking systems are essential to keep links stable without draining batteries.

\subsection{Underwater Networks}

\begin{figure*}
	\centering
	\includegraphics[clip, trim=0cm 5cm 0cm 0cm, width=0.9\linewidth, keepaspectratio]{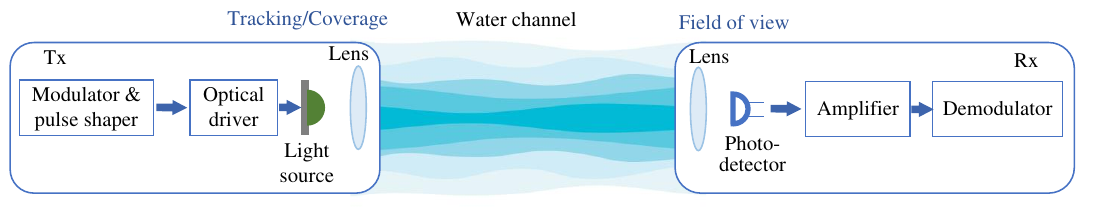}
	\caption{System model of an \acs{UWOC} system.}
	\label{fig:UWOC}
\end{figure*}

\begin{figure}
	\centering
	\includegraphics[width=0.75\linewidth]{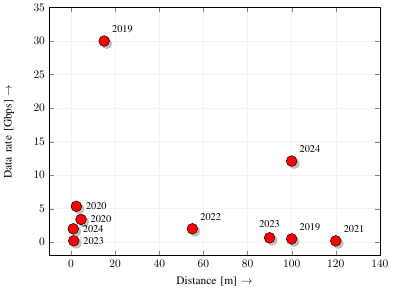}
	\caption{Advances in \acs{UWOC} over the past five years.}
	\label{fig:UWOCTTY}
\end{figure}

An emerging and promising area is \ac{UWOC}, which uses light waves to transmit information in underwater environments. 

\subsubsection{Opportunities}

Compared to traditional \ac{UEC} and \ac{UAC}, \ac{UWOC} offers several significant advantages such as high data rates, low latency, and high security. Recent research on \ac{UWOC} has focused on enhancing data rate and transmission distance, which have grown steadily over the past five years, see Fig.~\ref{fig:UWOCTTY}.

Careful selection of light sources and detectors is crucial for the success of \ac{UWOC} systems. \acp{LED} and \acp{LD} operating in the $470$--$535$ nm range, which avoids water absorption, are most suitable for short-range, medium-data-rate and long-range, high-data-rate operations, respectively. Recently, the use of alternative high-bandwidth sources such as micro-sized \acp{LED} and \acp{VCSEL} is also under consideration \cite{Soltani2023}. To enhance data rates and range, high-sensitivity and low-cost \acp{PMT} and \acp{MPPC} are deployed. The underwater environment is particularly suitable for these detectors due to its low background light.

\subsubsection{Challenges and Future Research Directions}
The unique characteristics of underwater environments, such as strong turbulence, scattering, and absorption, significantly affect communication performance of \ac{UWOC} systems. Promising solutions to overcome the effects and novel \ac{UWOC} research areas are outlined in the following.

\textbf{\ac{UWOC} networks:} \ac{UWOC} sensor networks are promising for future marine environmental monitoring and resource exploration. Recently, systems such as AquaE-net~\cite{UWOC_AquaFi}, a bi-directional LED-based \ac{UWSN}, are under research and development.

\textbf{Underwater \ac{SLIPT}:} \ac{SLIPT}, especially in combination with solar cells, is particularly important for overcoming the energy constraints of \ac{IoUT} devices. Enhancing the harvested energy while maintaining an acceptable system \ac{SNR} level is an ongoing research area for \ac{UWOC}--\ac{SLIPT}.

\textbf{\ac{RIS}-\ac{UWOC}:} \Ac{STAR}-\acp{RIS}, which have been studied in \ac{RF} systems for enhancing energy and spectral efficiency, have also been applied to \ac{UWOC} systems for overcoming turbulence and improving the luminous intensity at the receivers.

\subsection{Intra- and Inter-System Interconnecting Networks}
\label{sec:interconnects}

As computing demands grow, particularly in support of AI, big data, and cloud services, the performance of intra-system interconnects becomes increasingly critical. Faster, more scalable interconnects reduce bottlenecks, improve system responsiveness, and enable high-demand applications. Intra-system interconnects span from centimeters (onboard) up to tens of meters (rack-to-rack). For providing hundreds of individual links at speeds in excess of 10 \ac{Gbps} the use of conventional copper cables becomes increasingly difficult \cite{Cheng2024100Gbps}.

Furthermore, \acp{DCN}, which are poised to play a critical role in \ac{6G} networks for facilitating the storage and processing of massive data volumes, employ a hierarchical tree topology with racks housing dozens of servers and produce terabit-scale traffic. Handling such networks is difficult using fixed-capacity wired interconnects. Therefore, high data rate laser-based \ac{OWC} links have emerged as a scalable and cost-efficient approach for high-capacity interconnects between top-of-rack switches \cite{Celik2019DCN}.

Moreover, \acp{OI} are also increasingly gaining interest as an alternative to conventional copper links with high-speed \acp{OI} utilizing optical fibers, slab and dielectric waveguides, or \ac{FSO} links. Furthermore, \acp{OI} can be deployed at multiple levels, e.g., as board-to-board, chip-to-chip, or intra-chip interconnects.

\subsubsection{Opportunities}
The use of \ac{OI} offers several intrinsic advantages over electrical signal transmission, which suffers from physical limitations such as high noise, lower efficiency, and instability caused by the cable geometry. Furthermore, parasitic resistance, capacitance, and inductance restrict data transfer rates, especially at higher frequencies. On the other hand, optical cables offer reduced size and weight for similar applications.

\Acp{OI} can enhance overall data throughput as several optical wavelengths can be multiplexed on a single link, enabling complex computer applications such as AI processing and data analytics. In addition, with decreasing chip sizes, \acp{OI} offer improved scalability in both distance (centimeter-scale) and density by combining miniature light sources and photodetectors, and facilitate the creation of more complex and densely packed chips. We, therefore, see a strong increase in silicon photonic technologies.

Furthermore, as mentioned earlier, \acp{OI} can also offer scalable and cost-efficient high-capacity interconnects between top-of-rack switches.

\subsubsection{Challenges and Future Research Directions}
Despite their advantages, several areas of \ac{OI} remain challenging for widespread adoption and necessitate more research. These are outlined below.

\textbf{Thermal stabilization and fabrication imperfections:} Temperature variations severely impact the performance of optical components such as micro-ring resonators and lasers used in \ac{OI} systems and lead to wavelength shifts and power inefficiencies, necessitating robust thermal stabilization. Furthermore, the efficiency of light propagation and signal losses in the components are heavily dependent on the medium's effective refractive index and dispersion characteristics. Moreover, maintaining uniform waveguide dimensions across large-scale integrated photonic circuits is difficult, necessitating further research on advanced manufacturing techniques such as high-resolution lithography and etching processes.

\textbf{Power consumption and signal integrity:} Continuous laser operation and optical-to-electrical conversions in \acp{OI} result in high static and dynamic power consumption, which needs to be taken into account when designing \ac{OI} systems. Similarly, signal integrity may be compromised at waveguide junctions and bends, resulting in light leakage crosstalk and coupling losses. Self-sufficient receivers offer pathways to reduce power consumption. Using the optical channel itself to power the receiver using \ac{SLIPT} approaches is part of ongoing research. 

\textbf{Design automation tools:} Large-scale \ac{OI} systems need effective topologies to control latency and manage bandwidth. More research may be needed in designing dynamic reconfiguration strategies and hybrid architectures that combine electrical and optical networks, thereby ensuring that the full potential of low-power, high-bandwidth, and scalable \acp{OI} is achieved.

\subsection{Vehicle-to-Everything (V2X) Networks}

\begin{figure*}
	\centering
	\includegraphics[clip, trim=0cm 0cm 2.65cm 0cm, width=0.8\linewidth, keepaspectratio]{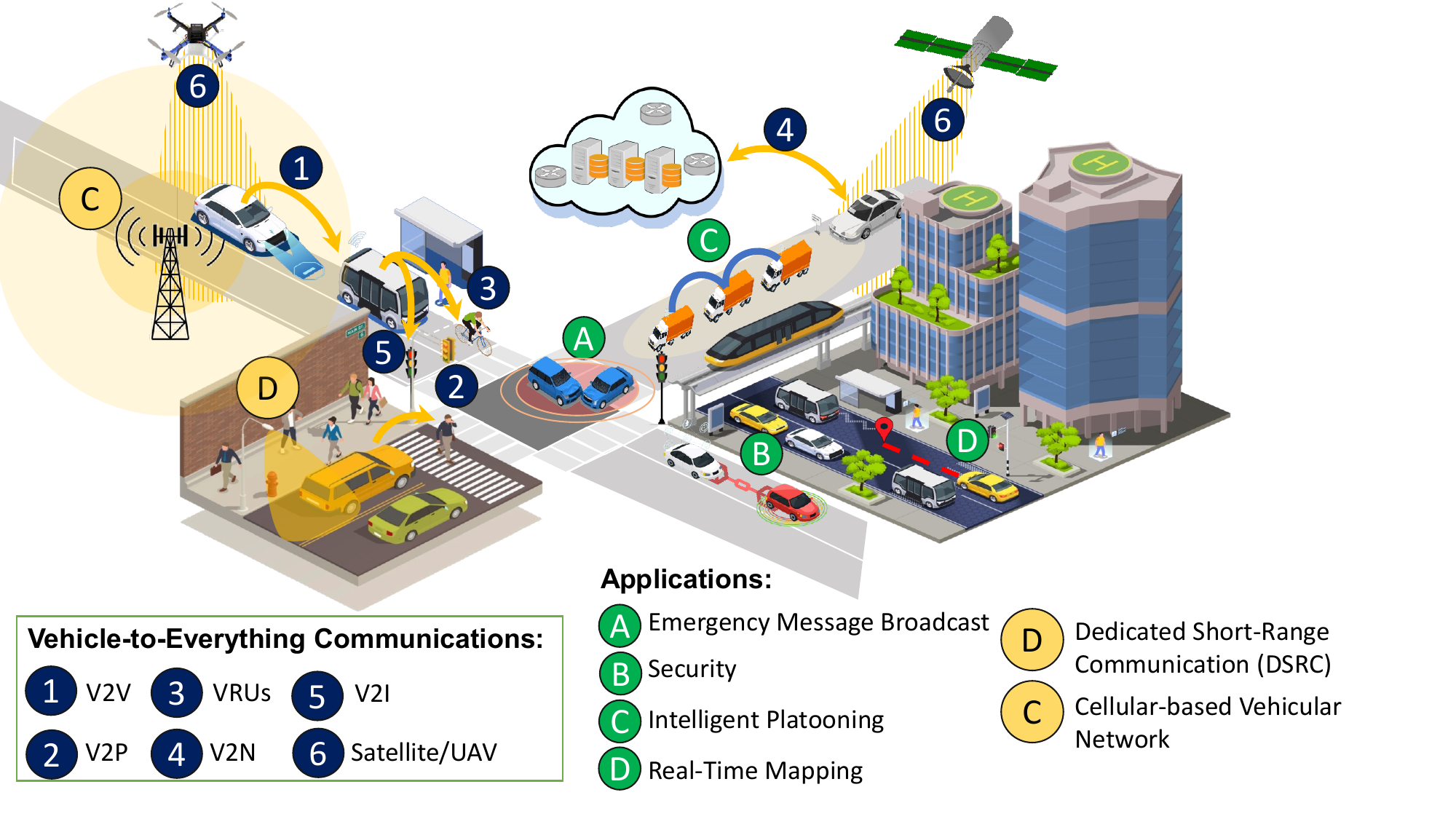}
	\caption{Integrated ITS system with various transportation scenarios and use cases. (Some of the vectors in the figure were designed by Freepik.)}
	\label{fig:V2X}
\end{figure*}

As we transition into an era of connected autonomous vehicles, the development of advanced \ac{V2X} communication links becomes key to realizing safe, efficient, and autonomous transportation networks. \ac{V2X} enables wireless communication between \ac{V2V}, \ac{V2I}, and \ac{V2P} communications, as well as communications with \acp{VRU} and cloud networks, i.e., \ac{V2N}. As a part of \acp{ITS}, \ac{V2X} has the potential to offer significantly enhanced road safety, air quality, and efficiency, leading to fewer traffic jams and an unprecedented user and application experience, as shown in Fig.~\ref{fig:V2X} \cite{Zhou2020}.

\subsubsection{Opportunities}
The integration of \ac{ITS} with \ac{V2X} can revolutionize transportation by enhancing safety, efficiency, and sustainability. \Ac{OWC} plays an important part in \ac{V2X} by enabling ultra-high-bandwidth data links for applications such as autonomous driving, real-time traffic monitoring and management. Moreover, \ac{OWC}'s robust and interference-free communication links can enable \ac{V2V} and \ac{V2I} applications.

\subsubsection{Challenges and Future Research Directions}
Although the use of \ac{OWC} for \ac{V2X} systems is promising, a number of challenges must be addressed before widespread deployment, which are outlined below.

\textbf{Limited transmitter field of view and precise beam alignment in high-mobility scenarios:} Optical beam tracking mechanisms to overcome vehicle vibrations and jerks are necessary. Furthermore, research on hybrid \ac{RF}-\ac{OWC} \ac{V2X} systems to mitigate the effects of adverse weather conditions, such as fog and rain, is critical.

\textbf{Standardized protocols:} The development of standardized protocols for integrating \ac{OWC} with the current \ac{RF}-based \ac{V2X} ecosystems is important.

\section{Emerging \acs{OWC} Technologies} \label{sec:emerging}

In this section, we discuss emerging \ac{OWC} technologies such as such as \acp{PV}, \ac{OAM}-based \ac{OWC}, \acp{ORIS}, \acp{OLED} and \acp{OPD}, and novel \ac{UV} devices that will enhance next generation \ac{OWC}-based \ac{NoN} across diverse domains.

\subsection{Photovoltaic Cells}

Addressing the energy and communication requirements of \ac{IoT} devices is crucial for maintaining seamless connectivity of vast and diverse next-generation \ac{IoT} networks. \ac{SLIPT} systems, which utilize photovoltaic cells with dual functionalities for energy harvesting and signal acquisition, have emerged as a viable solution for addressing the requirements. On the one hand, energy harvesting via photovoltaic cells overcomes energy constraints of \ac{IoT} devices, thereby enabling \ac{OWC} across satellite, aerial, terrestrial, and underwater domains. On the other hand, the use of photovoltaic cells as detectors allows for energy-efficient, environmentally friendly, and cost-effective devices compared to conventional detectors like \ac{PIN} diodes and \acp{APD} \cite{kong2020survey}.

Silicon-based devices are by far dominating the market for solar cells. The technology matured, with record lab cell efficiencies of $27.4\%$ \cite{green2024solar} approaching the theoretical efficiency limit of $29.4\%$. At the same time, technological advancements, economies of scale and standardization have led to impressive cost reduction, a learning rate for levelized cost of solar electricity of $38.5\%$ in recent years, and module prices well below $0.5~\$/W_p$ \cite{IRENA2024}. Besides, conventional single-junction silicon solar cells, \ac{MJ} designs, also known as tandem solar cells, are used to convert broad-band sunlight more efficiently. These devices are made of several vertically stacked subcells, each composed of a different absorber material. A record solar conversion efficiency of $47.6\%$ was achieved using a wafer-bonded four-junction concentrator solar cell under $665$ suns \cite{helmers2024advancing}.

\subsubsection{Opportunities}

Unlike solar cells that operate under the broad-band solar spectrum, \acp{PPC} are photovoltaic devices designed for operation under narrow-band or monochromatic light. Here, by tuning the absorber material's bandgap to match the photon energy, transmission and thermalization losses can be significantly minimized \cite{hohn2016optimal}. Advanced optical engineering further enables the use of resonances, leveraging micro-cavities and waveguide modes in ultra-thin devices \cite{hohn2022realization}. Furthermore, radiative recombination within the \ac{PPC} can produce emitted light that can be trapped and subsequently reabsorbed, known as photon recycling, which enhances the internal carrier density thus increasing the voltage in a manner similar to the effects of concentration \cite{6213058}. Using a thin-film processing approach with backside reflector, optical resonance, and photon recycling have been utilized to achieve a \ac{GaAs}-based \ac{PPC} with a record \ac{PCE} of $68.9\%$ under $858$~nm monochromatic light \cite{helmers202168}. Unlike conventional solar cells, which have areas ranging from $24.4$~cm${}^2$ (M0 format) to $44.1$~cm${}^2$ (M12 format), \ac{PPC}s are much smaller, typically ranging from a few square-millimeters to 1 cm${}^2.$ This smaller size enables much higher bandwidths when using a \ac{PPC} as a receiver.

\begin{figure}
	\centering
	\includegraphics[width=0.9\linewidth]{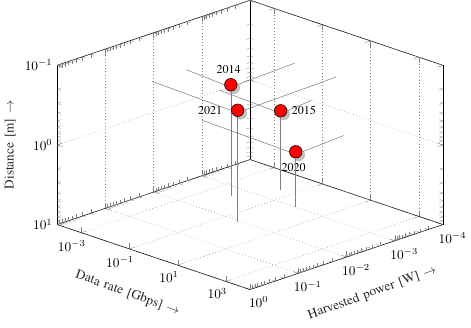}
	\caption{Schematic showing the distance, data-rate, and harvested power achieved over the years.}
    \label{fig:Solarcell}
\end{figure}

First experimental studies have been conducted on \acp{SLIPT} using photovoltaic receivers, where modulated laser light was used to carry power and data from a transmitter to the \ac{PPC} receiver. The state-of-the-art advancements in this technology, considering data rates and energy harvesting over varying distances, are illustrated in Fig.~\ref{fig:Solarcell}. In \cite{fakidis2020simultaneous}, operating a \ac{GaAs}-based \ac{PPC} under eye-safe low-power laser illumination across a 2-m free-space link, a \ac{PCE} of $41.7\%$ was demonstrated with power harvesting of $1$~mW and simultaneous data transfer at a rate of $0.78$~\ac{Gbps}. Furthermore, by adjusting the operating point away from the \ac{MPP} and moving closer to short-circuit conditions, even higher data rates of $1.04$~\ac{Gbps} were achieved.

Hence, \ac{SLIPT} via \acp{PPC} has a tremendous potential to transform next generation \acp{NoN} from the energy harvesting and communication perspectives.

\subsubsection{Challenges and Future Research Directions}

Despite advancements in photovoltaic technology, several challenges remain in the development of high-performance \ac{SLIPT} systems. The key challenges and potential research directions are outlined below.

\textbf{Absorber material:}
High quality absorber materials with band-gap energies well matched to the photon energy of incident light are crucial to obtain high conversion efficiencies. By tuning the composition of ternary or quaternary III-V compound semiconductors, a wide spectral range from the visible to near-infrared light is accessible \cite{helmers2024advancing}.
 
\textbf{\ac{MJ} device architectures:} 
\ac{MJ} architectures, analogous to those in high-efficiency solar cells, are also possible for \acp{PPC}. \ac{MJ} \acp{PPC} are vertical stacks of multiple subcells made of the same absorber material. Therefore, their response is spectrally highly selective and the subcell thicknesses must be carefully designed to obtain current matching at the transmitter's wavelength. Thus, all subcells receive the same signal, allowing direct extraction at external terminals. Furthermore, the series connection of the subcells increases the output voltage while reducing the overall capacitance, promising devices with high communication bandwidths.

\textbf{Multi-segment device architectures:} 
Similar to \ac{MJ} devices, multi-segment \ac{PPC}s are on-chip modules consisting of a series-connection of several subcells \cite{kimovec2019comprehensive}. However, unlike \ac{MJ} \acp{PPC}, individual segments are interconnected laterally. This way, the devices benefit from both the reduced capacitance and increased voltage output from the integrated series-connection, while featuring a broad spectral response. However, lateral transport typically comes at the cost of increased series resistance and cutbacks in conversion efficiency.

\textbf{Bandwidth--harvested power device trade-off:} 
Smart \ac{SLIPT} device architectures that can optimally address the trade-off between bandwidth and total harvested power are necessary. In this regard, \ac{MJ} or multi-segment \acp{PPC} described above are of interest as they can reduce capacitance while maintaining a large photosensitive area.

\textbf{Rate-power operational trade-off:} 
In practical \ac{SLIPT} systems, managing the trade-off between data rate and power harvesting efficiency is as well a critical challenge. In dynamic systems, requirements can change over time, making it important to explore techniques for allocating adequate data rates and power. In this context, the addition of varying DC bias light to the alternating current (AC) signal, along with balancing the receiver's operating point between the maximum data rate in short-circuit conditions and the maximum power point, is of particular interest.

\textbf{MIMO:}
The use of \ac{MIMO} systems may be beneficial to improve both resilience and efficiency in \ac{SLIPT} systems. To this end, the authors of \cite{de2024reconfigurable} introduce a \ac{MIMO} system design which achieves a data rate of $85.2$~Mbps while simultaneously harvesting energy from sunlight, with a maximum harvested power of $87.33$~mW.

\subsection{OAM}

\Ac{OAM}-based \ac{OWC} is emerging as a novel method to support high data rates. Here, data is encoded into multiple concurrent \ac{OAM} states of light, allowing a single optical beam to carry several independent data channels simultaneously, thereby enabling efficient data multiplexing.

\subsubsection{Opportunities}

An \ac{OAM} state of a light beam exhibits a helical phase structure and is mathematically represented as $\exp(i l \phi)$, where the topological charge $l$ and the azimuthal angle $\phi$ can be adapted to encode digital data. Beams with different topological charges are orthogonal to each other. Furthermore, they are robust against crosstalk and beam perturbation, and are secure against eavesdropping. Hence, exploiting \ac{OAM} states of light is ideal for enabling high-data-rate \ac{OWC}. In fact, \ac{OAM} with a single wavelength or polarization can already significantly increase data transmission capacity without extra bandwidth requirements \cite{wang2012terabit}. Furthermore, foundational work in \cite{willner2016design} established the feasibility of \ac{OAM} for high-capacity communication in both free-space and fiber-optic networks.

\begin{figure*}
	\centering
	\includegraphics[clip, trim=2cm 3.4cm 2cm 0cm, width=0.8\linewidth]{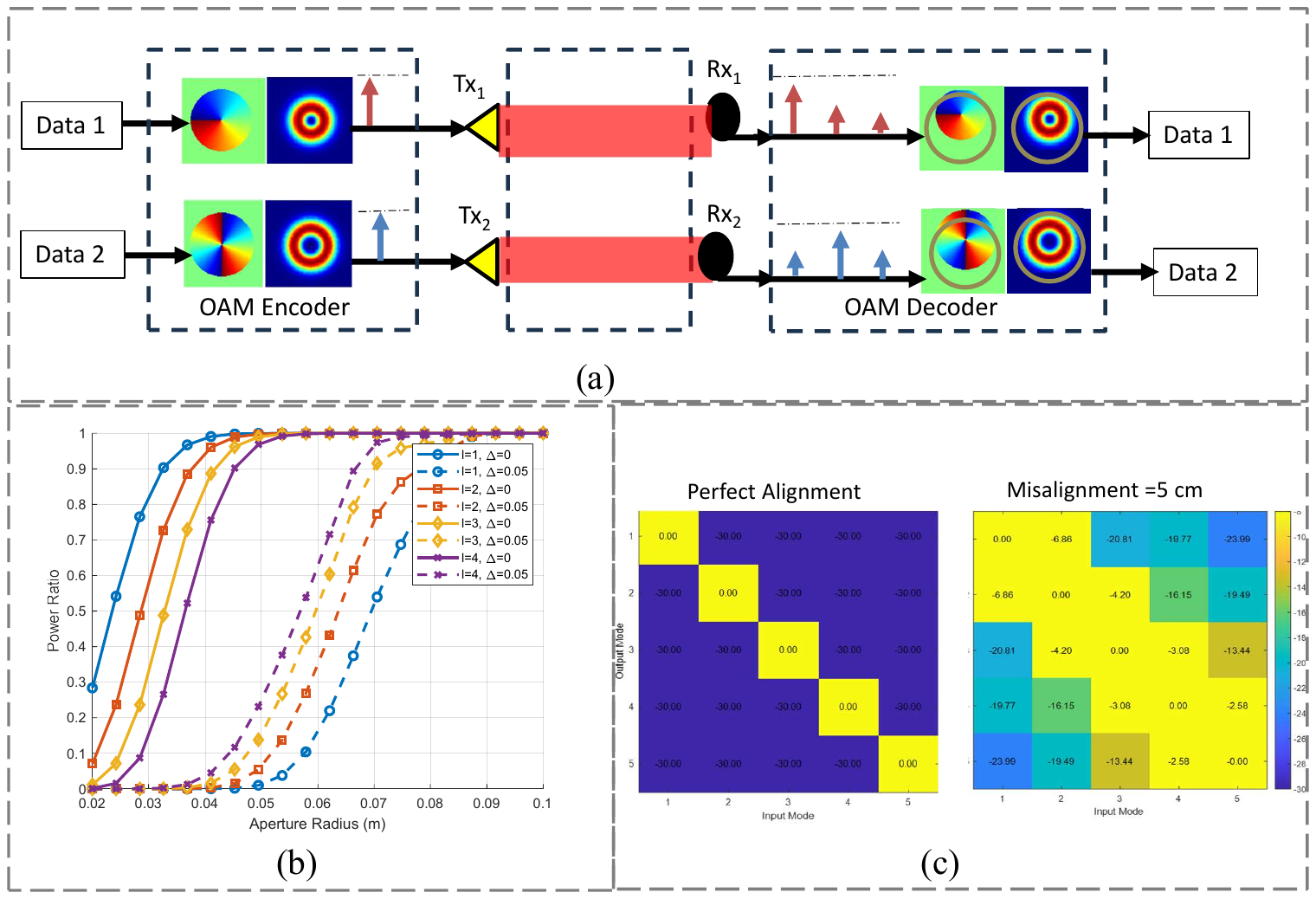}
	\caption{(a) Block diagram of an \ac{OAM}-based \ac{OWC} system employing two modes, (b) power ratio vs. aperture radius for aligned and misaligned ($\Delta = 0.05$~m) configurations, and (c) matrix showing interference between modes for aligned and misaligned configurations.}
	\label{fig:OAM}
\end{figure*}

Next, we discuss the impact of misalignment on \ac{OAM}-based \ac{OWC}. As shown in Fig. \ref{fig:OAM} (a), the total channel capacity increases with the number of users.  However, misalignments, power leakage, and inter-modal interference may occur, which leads to a significant decrease in capacity. As shown in Fig.~\ref{fig:OAM} (b), for perfectly aligned systems, even a small receiver aperture ($6$ cm) efficiently captures nearly all power, even for higher-order modes ($l$ = 4). However, even with slight misalignment, the received power drops significantly—down to $60\%$ for $l$ = 4. Fig.~\ref{fig:OAM} (c) illustrates cross-talk levels for both aligned and misaligned scenarios. With perfect alignment, cross-talk is minimal. However, in the presence of misalignment, the cross-talk increases and the neighboring modes have comparable power levels. Nevertheless, despite this increase in inter-modal interference, cross-talk still remains mode-dependent. That is, for users with modes $l = 1$ and $l = 5$, cross-talk is limited to $23$ dB below the intended signal level, even with misalignment. Hence, careful selection of modes has the potential of mitigating cross-talk effects due to misalignment.

\subsubsection{Challenges and Future Research Directions}
Looking forward, \ac{OAM}-based \ac{OWC} systems hold significant promise for high-capacity, efficient data transmission.

\textbf{Metasurfaces:} Additionally, recent advancements in metasurface technology enable the generation of \ac{OAM} beams with ultrathin structures, achieving high conversion efficiency across broad bandwidths and independent mode generation for different polarizations. Furthermore, \ac{OAM} emitters with on-demand features, such as collimation, controllable directionality, and multi-channel output, can now be realized via monolithic integration of metasurfaces with \acp{VCSEL}. However, currently, only passive metasurfaces with fixed electromagnetic responses have been utilized for optoelectronic integrations, limiting their use for more advanced applications such as \ac{OAM}-based reconfigurable multiplexing and adaptive tuning.

\textbf{Algorithms:} Future research should prioritize computationally efficient algorithms for optimal mode selection and power allocation. In this regard, the use of \ac{AI}/\ac{ML} for dynamic \ac{OAM} modes selection is promising for enabling real-world deployments of \ac{OAM}.

\subsection{Optical RIS} \label{sec:oris}

\Acp{RIS} have emerged as a transformative solution that enable dynamic manipulation of optical signals, including beam direction, phase, and amplitude, for facilitating high-speed, large-scale, multi-user \ac{OWC} \cite{Najafi2019RIS,DigitalRIS}.

\begin{figure}
    \centering
    \includegraphics[width=0.9\linewidth]{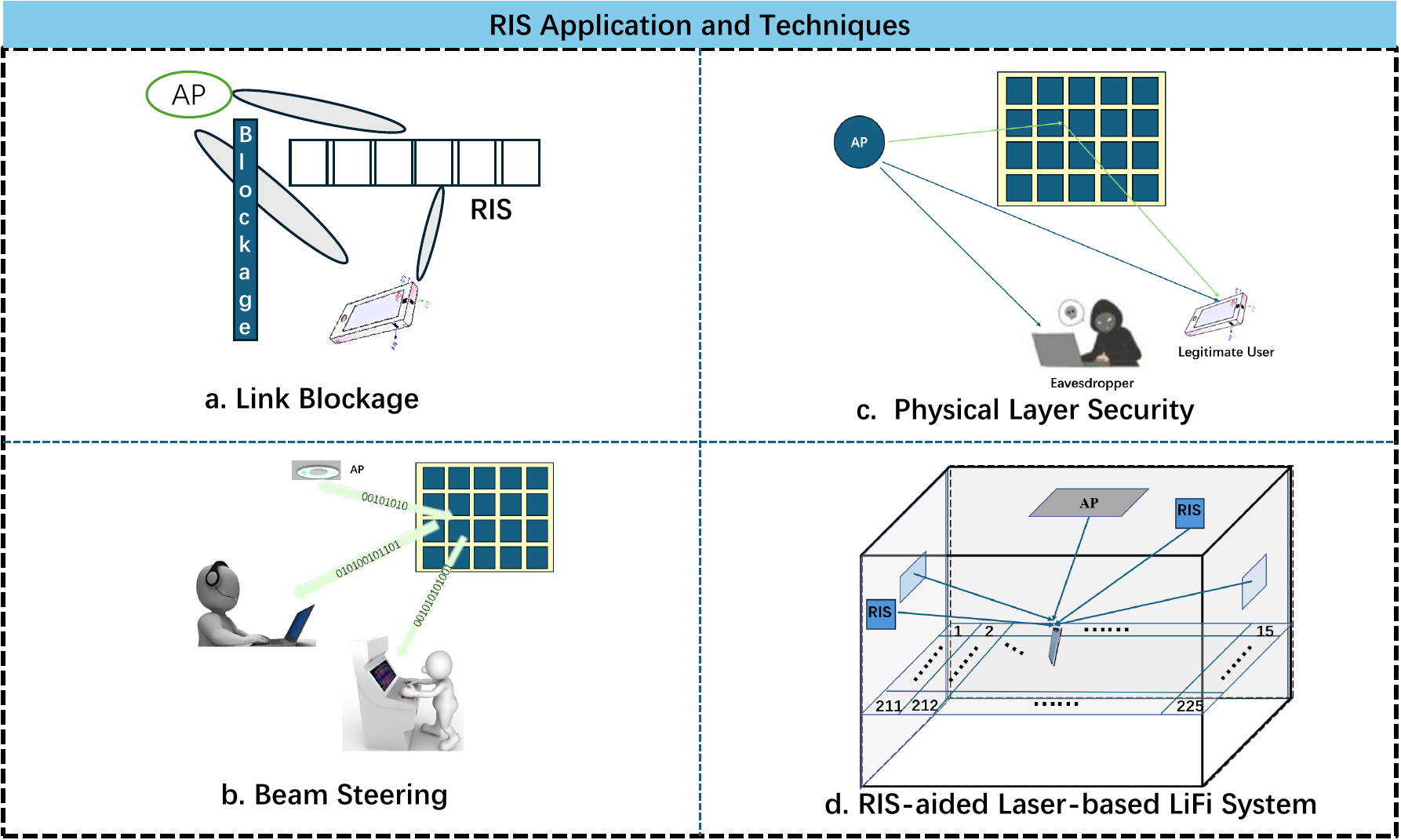}
    \caption{Potential applications for \ac{RIS}-aided \ac{OWC} systems include link blockage avoidance, light beam steering, physical layer security, and augmented \ac{RIS}-\ac{OWC} systems such as \ac{RIS}-aided \ac{LiFi} systems.}
    \label{fig:RIS}
\end{figure}

\Acp{RIS} are engineered surfaces capable of dynamically controlling the reflection, phase, amplitude, and polarization of incident electromagnetic waves. They consist of arrays of programmable optical elements, such as micro-mirrors or metasurfaces, which, unlike traditional reflective surfaces, can precisely manipulate light beams, to adapt to the environmental conditions and user demands. Selected potential applications of \ac{RIS} in \ac{6G} and beyond \ac{OWC} systems are shown in Fig. \ref{fig:RIS}.

\subsubsection{Opportunities}
\ac{RIS} effectively addresses link blockages in \ac{OWC} systems by reconfiguring the direction and phase of optical signals to establish alternative \ac{NLOS} paths around obstacles, thus maintaining reliable connectivity, as shown in Fig. \ref{fig:RIS} (a).

In fact, as shown in \cite{maraqa2023optimized} for multi-user scenarios, joint mirror arrays and \ac{LC}-based \ac{RIS} system can achieve $196\%$ improvement in sum rate over conventional systems.
\Ac{RIS} can also enhance physical layer security, as shown in Fig. \ref{fig:RIS} (b), by precisely aligning reflective elements and confine signal propagation to designated areas, thus reducing the risk of eavesdropping.

Furthermore, as shown in Figs \ref{fig:RIS} (c) and (d), integrating \ac{RIS} into \ac{OWC} systems can facilitate advanced signal processing techniques such as beam steering and overcome limitations of systems such as \ac{LED}-based \ac{LiFi} systems.

\subsubsection{Challenges and Future Research Directions}
In the following, we outline the key challenges in designing optical \acp{RIS}.

\textbf{\ac{RIS} placement and power allocation:} Despite the above advantages, the placement of \acp{RIS} in \ac{OWC} systems and careful power allocation are crucial to exploit the full potential of \ac{RIS}-aided \ac{OWC} systems, see, e.g., \cite{huang2024energy}.

\textbf{Realtime \ac{RIS} configuration:} Hardware and software support for realtime configuration of the \acp{RIS} is essential to dynamically control the reflection of the incident electromagnetic waves. The use of \ac{AI}/ac{ML} algorithms for realtime control is also of interest.

\subsection{OLED/OPD}

\begin{figure}
	\centering
	\includegraphics[width=0.75\linewidth]{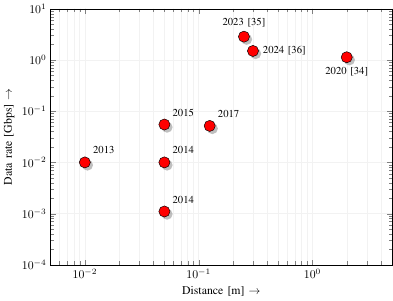}
	\caption{Measured data rates as a function of transmission distance from previous works, evaluated at a reference BER of $3.8\cdot 10^{-3}$.}
	\label{fig:OLED}
\end{figure}

\Acp{OLED} and \acp{OPD} are cost-effective alternatives to conventional \acp{LED} and \acp{PD} and have the potential to enable a wide range of novel applications.

\subsubsection{Opportunities}

\Acp{OLED} are a promising alternative to conventional \acp{LED} owing to their low cost, mechanical flexibility, and the ability to integrate multiple \ac{OLED} pixel colors on a single substrate, and are anticipated to enable diverse paradigms from display-to-display communication to \ac{VLC}. \Acp{OLED} have been in active research and development in the last two decades, making significant progress in terms of communication distance as well as data rates, see Fig.~\ref{fig:OLED}. Furthermore, the demands of high-speed applications, recent advances have significantly enhanced the response times of \acp{OLED}  \cite{10308568}. In fact, a data transmission rate of $3.2$ Gbps over a $30$ cm link with a \ac{RGB}-\ac{OLED} pixels was recently achieved \cite{yoshida2024rgb}.

On the other hand, \acp{OPD} are organic alternatives to \acp{PD}, which, in addition to photo detection can incorporate optical filters, built-in lenses, or other optical components to enhance their performance. Furthermore, unlike traditional \acp{PD}, \acp{OPD} exhibit superior responsivity in the visible range, making them particularly advantageous for \ac{VLC} systems. Furthermore, \acp{OPD} can be manufactured as ultrathin photoactive layers of $100$~nm thickness. Moreover, high-speed \acp{OPD} with data transmission rate of $80$~Mbps have been demonstrated \cite{zhu2024high}, thus enabling cost-effective, flexible, and energy-efficient receivers for next-generation integrated and scalable \ac{LiFi} modules.

\subsubsection{Challenges and Future Research Directions} 
The design and development of \acp{OLED} and \acp{OPD} face numerous challenges, primarily related to material limitations, device performance, and environmental impact. Below, the key issues affecting both \acp{OLED} and \acp{OPD} are summarized \cite{liguori2024overcoming, zhang2023organic}.

\textbf{Reliability and stability:}
Both OLEDs and OPDs degrade in performance over time due to electrical stress, trap density, humidity, and oxygen exposure. \Acp{OLED}, especially larger panels, experience faster degradation and color fading. As well, \acp{OPD} face challenges in stability when exposed to environmental factors like moisture and oxygen. Addressing these issues requires improved material engineering, passivation techniques, and enhanced design to ensure better longevity and performance.

\textbf{Material limitations:}
\Acp{OLED}, particularly white \acp{OLED}, rely on complex techniques such as thermally activated delayed fluorescence to achieve high efficiency, which requires intricate molecular designs and poses development challenges. On the other hand, \acp{OPD} also suffer from material limitations, including poor charge carrier mobility and reduced thermal and mechanical stability. Further development of organic semiconductors with improved properties is essential for enhancing both \ac{OLED} and \ac{OPD} performance.

\textbf{Performance limitations:}
For \acp{OLED}, challenges include low luminescence on small surfaces, emission uniformity, and maintaining transparency in contacts. \Acp{OPD}, on the other hand, face performance challenges like high dark current density, low external quantum efficiency, and noise issues such as shot and $1/f$ noise. These performance limitations reduce sensitivity and overall efficiency, especially in low-light conditions or high-frequency applications. %

\textbf{Response speed and dynamic range:}
\Acp{OPD} exhibit slow response speeds, characterized by low cutoff frequencies and high rise and fall times, restricting their suitability for high-speed optical communication systems. Similarly, \acp{OLED} face challenges in achieving uniform light emission over large areas and maintaining performance consistency. Both technologies need advancements in charge carrier mobility, material design, and device architecture to improve their response speed and dynamic range for broader and faster applications.

\subsection{Ultraviolet Communication}

\ac{UV} communication began in the twentieth century and has continually gained traction due to its unique ability among optical wavelengths to transmit data effectively in \ac{NLOS} scenarios \cite{Tian2022}.

\ac{UV} light may be classified according to wavelength into three categories, \ac{UV}-A ($315$ -- $400$ nm), \ac{UV}-B ($280$ -- $315$ nm) and \ac{UV}-C ($200$ -- $280$ nm), with the latter being the solar-blind region where all solar \ac{UV}-C radiation is absorbed by earth's upper atmosphere and virtually no \ac{UV}-C solar radiation is present at ground level.

\subsubsection{Opportunities}%

\ac{UV} offers unique and useful properties for \ac{OWC} compared to the visible--\ac{IR} spectrum. Firstly, Rayleigh scattering of light by the atmosphere is enhanced at short wavelengths, which can be exploited to support non-line-of-sight \ac{OWC}, where scattered light provides an indirect path between transmitter and receiver in case of obstructions. Furthermore, communication in the solar-blind region within the \ac{UV}-C band has minimal background interference.

In recent years, there have been significant improvements in the efficiencies, output powers, and range of coverage of the \ac{UV} spectrum of Aluminium Indium Gallium Nitride (AInlGaN) \acp{LED}, driven largely by the use of \ac{UV} \acp{LED} for sterilization of surfaces and water. %
Hence, this mature \ac{UV} \ac{LED} ecosystem is ripe for exploitation. In addition, \ac{UV} \acp{LED} offer improved size, weight and power properties compared to conventional \ac{UV} sources, such as mercury vapor lamps, and have far superior modulation bandwidths, which is crucial for \ac{OWC}.

\Ac{LOS} \ac{OWC} demonstrations using \ac{UV}-C \acp{LED} and micro-\acp{LED} have been reported by several research groups for transmission distances from a few tens of centimeters to several meters, with corresponding data rates ranging from many tens of \ac{Mbps} to several \ac{Gbps} \cite{Tian2022}. More recently, the use of high bandwidth \ac{UV} micro-\acp{LED} for $10$ \ac{Gbps} \ac{WDM} using \ac{UV}-A, -B, and -C emitters \cite{Maclure2022a}, as well as \ac{Gbps} \ac{OWC} at ranges exceeding $100$~m using a single \ac{UV}-C micro-\ac{LED} \cite{Maclure2022b} have been demonstrated.

\begin{figure}
    \centering
    \includegraphics[clip, trim=0cm 0.55cm 9.75cm 0cm, width=0.8\linewidth,keepaspectratio]{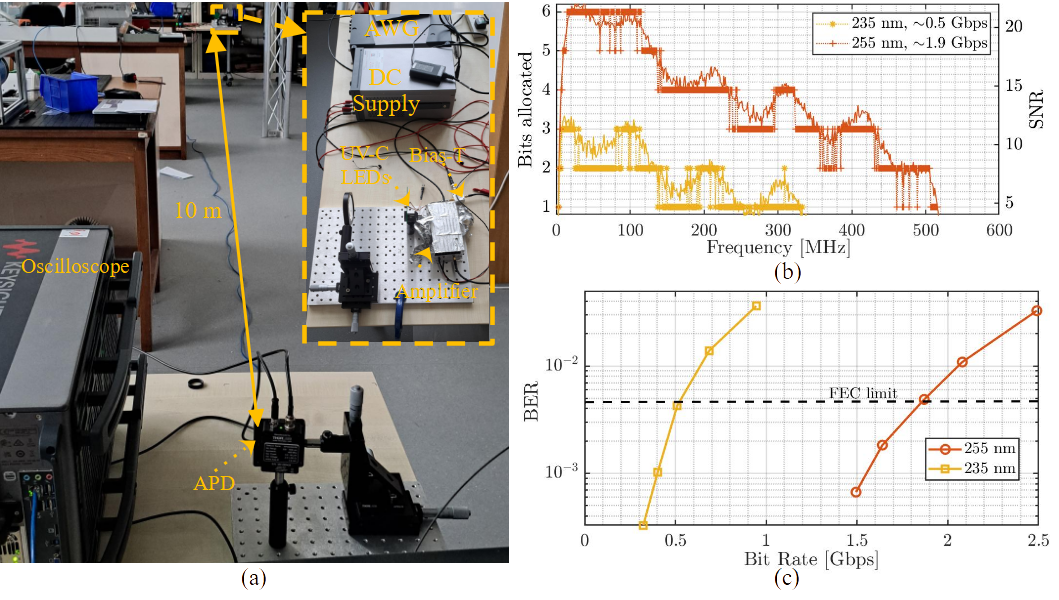}
    \caption{Experiment setup for a 10~m \ac{OWC} link  with \ac{UV}-C \acp{LED} (235~nm and 255~nm).}
    \label{fig:UVC10m1}
\end{figure}

\begin{figure}
	\centering
	\includegraphics[width=0.8\linewidth]{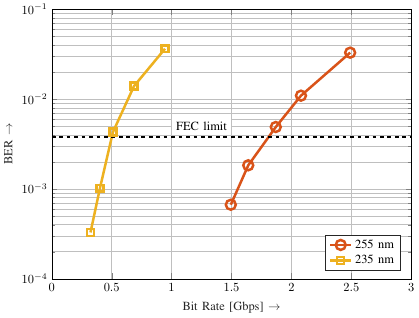}
	\caption{\ac{BER} vs. bit rate for 235 and 255~nm LEDs, respectively, at a line-of-sight transmission distance of 10~m using adaptive bit and power-loaded \ac{OFDM}.}
	\label{fig:UVC10m3}
\end{figure}

At $30$~cm link length, the work in \cite{zimi2023gigabit} reports bit rates of about $0.68$~\ac{Gbps} and $1.53$~\ac{Gbps} from the $235$~nm and $255$~nm LEDs, respectively. As shown in Figs.~\ref{fig:UVC10m1} and \ref{fig:UVC10m3}, more recently, the range has been extended to $10$~m with bit rates of about $0.5$~\ac{Gbps} and $1.9$~\ac{Gbps} from the $235$~nm and $255$~nm \acp{LED}, respectively.

\subsubsection{Challenges and Future Research Directions}
In the following, we outline the key challenges in \ac{UV} communication.

\textbf{Shorter wavelengths:} Thus far, \ac{UV} \ac{OWC} has most commonly been explored at wavelengths around $280$~nm, with limited research at shorter wavelengths. Research on shorter wavelengths would be advantageous both for operating in the true solar-blind region of the spectrum and skin and eye safety from \ac{UV} exposure. As such, exploring the use of \ac{UV}-C LEDs with peak emissions at $235$ and $255$~nm is advantageous \cite{Nicholls2023}.

\textbf{Link length:} Research into extending the link length of \ac{UV} \ac{LOS} links is of high interest for developing practical \ac{UV} communication systems.

\section{Value-Added Technologies} \label{sec:6gtech}

In this section, indoor positioning and gesture recognition technologies, which capitalize on \ac{OWC} integrated networks, are presented.

\subsection{Indoor Positioning} \label{sec:indoorpositioning}

\begin{table*}
	\centering
	\caption{Summary of Recent Works towards High Precision LiFi Positioning. E/S: Experimental/Simulation study. Accu: Accuracy.}
	{\footnotesize
		\renewcommand{\arraystretch}{1.5}
		\begin{tabular}{| l  r  p{3.5cm}  r  p{6cm} |}
			\hline
			\textbf{Reference} & \textbf{Accu. (cm)} & \textbf{Technique} & \textbf{E/S} & \textbf{Remark}\\\hline\hline
			Bakar \emph{et al.}, 2020 \cite{bakar2020accurate}
			 & $0.5$ &  Fingerprinting + \ac{DNN} & E & With \ac{2D} estimation, multiple \acp{Rx}, area $400 \times 400$ mm \\
			He \emph{et al.}, 2019 \cite{he2019demonstration}
			 &  $0.9$ &  With \ac{RSS} based triangulation using \ac{ANN}  & E & Multiple \acp{Tx}, area $900 \times 900$ mm \\
			Shao \emph{et al.}, 2018 %
			 & $2$ & \ac{RSS} + triangulation & E & Multiple \ac{Rx} and retroreflector \\
			Huang \emph{et al.}, 2024 %
			 & $2.31$ &  Fingerprinting + \ac{CNN} & S &  Uses \ac{ADR} and angle correction for joint \ac{3D} position and orientation estimation. \\
			Jeon \emph{et al.}, 2024 %
			 & $3$  & Fingerprinting + ToA + \ac{CNN} & S & Joint \ac{3D} position and orientation estimation using deep \ac{RSS}-\ac{ToA} fusion network\\
			Ma \emph{et al.}, 2022 %
			 & $3.3$  & \ac{ToA} + reference signal & E & Requires three \acp{Tx}, uses angle correction to reduce error from $5.3$ to $3.3$ cm \\
			Kokdogan \emph{et al.}, 2024 %
			 & $4.2$ & RSS-based \ac{MLE} & S & Adjusts the orientation of \ac{RIS} based on $N$-step localization algorithm to improve the accuracy \\
			Ahmad \emph{et al.}, 2023 %
			 & $4.5$  & Fingerprinting + \ac{DNN} & S & Joint \ac{3D} position and orientation estimation. Array of \acp{VCSEL} as transmitter \\
			Kouhini \emph{et al.}, 2021 %
			 & $5$ & ToA & E & \ac{2D} estimation with simultaneous \ac{Tx} and \ac{Rx} rotation \\
			De Bruycker \emph{et al.}, 2024 %
			 & $5.3$ & TDoA & E & Area $500\times500$ mm$^2$ \\
			\hline
		\end{tabular}
	}
	\label{tab:reltdwork-pos}
\end{table*}

Indoor positioning systems are critical for enabling a wide range of emerging applications for \ac{6G} technology, including precise navigation, real-time tracking, and automation within enclosed spaces.
Conventional \ac{GPS} is inaccurate and imprecise for indoor positioning due to the significant attenuation of satellite signals caused by building infrastructure. In contrast, current indoor positioning systems, such as \ac{WiFi} and Bluetooth, have limited accuracy due to signal shadowing and multipath propagation, typically ranging from $1$--$7$ m and $2$--$5$ m~\cite{ahmad2023joint}, respectively.  However, these accuracy levels are insufficient for next-generation high-precision applications, such as telesurgery and \ac{C2X} connectivity, which demand sub-centimeter accuracy.

\subsubsection{Opportunities}
Fortunately, \ac{OWC} systems are well suited for precise positioning. \Ac{LiFi}, unlike \ac{RF}-based systems, benefits from the widespread availability of light sources, absence of fading effects, and better resolution owing to small wavelengths, making LiFi-based positioning ideal for indoor applications and achieving centimeter-level precision. In fact, \ac{LiFi} positioning systems have already been deployed for enhancing visitor navigation at the Grand Curtius Museum in Belgium, support product localization at Aswaaq and E. Leclerc stores in Dubai and France, respectively, and facilitate personalized marketing at Carrefour Lille stores in France.

While traditionally, positioning was limited to \ac{2D}, a recent study~\cite{ahmad2023joint} has laid the groundwork for joint estimation of \ac{3D} position and orientation, where \ac{VCSEL}-based transmitters combined with fingerprinting and \ac{DNN} achieved a $4.5$ cm localization accuracy. A summary of recent notable contributions to high-precision LiFi-based localization is provided in Table~\ref{tab:reltdwork-pos}.

\subsubsection{Challenges and Future Research Directions}
In the following, we outline the key challenges in \ac{OWC}-based indoor positioning.

\textbf{\ac{LOS} obstruction:} A challenges in \ac{LiFi}-based positioning is overcoming \ac{LOS} obstruction. For this, \acp{RIS} can facilitate positioning despite absence of \ac{LOS}, see Section \ref{sec:oris}. Recent research with a refocusing \ac{RIS} has demonstrated a localization accuracy of $4.2$ cm for \ac{RIS}-assisted \ac{LiFi}-based indoor positioning \cite{kokdogan2024intelligent,he2019demonstration}.

\textbf{\ac{AI}/\ac{ML}:} Technologies such as \ac{AI} and \ac{ML} will play a crucial rule in enhancing the localization precision. While limited in test area, \cite{bakar2020accurate} have demonstrated sub-cm accuracy in a \ac{LiFi} positioning system, validating the claims of high-precision \ac{OWC}-based localization. Nevertheless, further research on \ac{AI}/\ac{ML} algorithms and demonstrations on larger test areas are necessary.

\subsection{Gesture Recognition}

\begin{figure}
	\centering
	\includegraphics[clip, trim=0cm 0cm 17.85cm 0cm, width=0.75\linewidth]{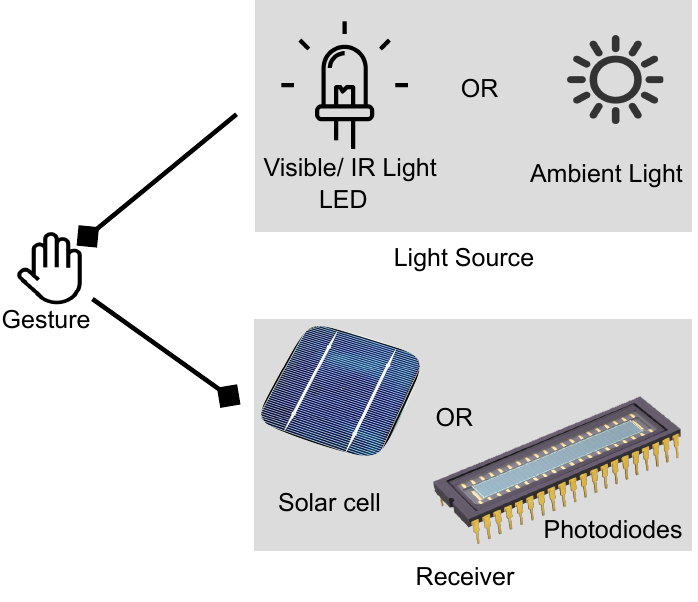}
	\caption{\ac{OWC} based gesture recognition can be based on dedicated or ambient light. Detection can be performed using solar cells or photodiodes.}
	\label{fig:Gesture Recognition}
\end{figure}

\Ac{OWC} systems enable non-contact gesture recognition by detecting the alterations to light signals due to gestures, which is beneficial for wearables, industrial automation, and \ac{IoT} applications. As shown in Fig.~\ref{fig:Gesture Recognition}, gesture recognition can be performed based on either dedicated or ambient light sources to suit specific environmental and operational needs. Furthermore, detection can also include dedicated gesture recognition devices or general devices such as solar cells.

\subsubsection{Opportunities}
Visible light and IR \acp{LED} can be modulated, allowing very precise control over the light signal which helps the system to differentiate between intentional gestures from background noise. Furthermore, \ac{IR} \acp{LED} provide an invisible light source, ideal for gesture recognition under low-light or fluctuating light conditions. Moreover, ambient sunlight or indoor lighting can be utilized in energy-constrained devices with energy harvesters to enable both energy harvesting and gesture sensing. For these systems, lightweight intelligent algorithms such as \ac{CFAR} \cite{Gesture3} are particularly useful which can dynamically adjust the detection threshold to account for changing light conditions and thus provide accurate gesture recognition.

On the other hand, for capturing gestures, arrays of photodiodes or solar cells are commonly used in \ac{OWC}-based gesture recognition systems. The work in \cite{Gesture1} has used an array of photodiodes in photoconductive mode, which recognize gestures with high sensing responsivity at the expense of external power input. In \cite{Gesture3}, photodiodes were employed in photovoltaic mode, to passively harvest energy from the ambient light and power the system and designed self-sustaining gesture recognition using ambient light. %

\subsubsection{Challenges and Future Research Directions}
There are many challenges in \ac{OWC}-based gesture recognition systems despite significant advancements.

\textbf{Algorithms:} Beyond hardware, signal processing and machine learning play an important role in the development of \ac{OWC}-based gesture recognition, especially for robust recognition under various lighting and gesturing conditions.

\textbf{Ambient light interference:} \ac{LED}-based systems, especially those using visible light \acp{LED}, can suffer from ambient light interference. Variations in sunlight or artificial lighting can lower the accuracy of gesture recognition considerably. For such systems, lightweight intelligent algorithms such as the \ac{CFAR} algorithm \cite{Gesture3} are necessary.

\textbf{Energy efficiency:} For wearable devices, research on lightweight data-processing algorithms and efficient photodiode arrays for reducing power consumption while maintaining the accuracy of gesture recognition are crucial.

\section{Conclusion} \label{sec:conclusion}

Metcalfe's law states that the value of a network scales quadratically with the number of connections. This is one of the key fundamental drivers for \ac{6G} and beyond. With the rapidly increasing number of autonomous systems, digital twins, and the cyber-physical continuum in general, future networks are set to expand terrestrially, into space, and underwater. 

Therefore, we cannot approach cellular networks in the same way as in \ac{4G} and \ac{5G} which primarily was a \emph{terrestrial} play. There is both a macro-scale (\ac{NTN}) and micro-scale (interconnects) expansion of networks, enabling a plethora of new use cases across multiple industrial verticals. As the innovation cycles within these verticals operate on different timescales, it will become increasingly challenging to maintain a unified clock cycle in the form of \emph{generations}. A fundamental rethinking of cellular network innovation cycles may be necessary, though this is beyond the scope of this paper.

This paper highlights the significant opportunity to deliver the ultimate network of networks by integrating optical wireless technologies alongside traditional \ac{RF} technologies. This approach enables spectrum to be utilized in a more targeted and efficient manner, accommodating the diverse connectivity requirements within a coordinated and converged \ac{NoN}.

In this context, it may be useful to define a new metric, \emph{usefulness per Hertz bandwidth}, as it is not practical to rely on \ac{RF} for long-distance underwater communication or to \emph{force} RF signals through highly attenuating walls and windows in a bid to use conventional mobile networks indoors. Instead, this paper argues that we should leverage all available tools and spectrum to develop an intelligent future \ac{NoN}.

\bibliographystyle{IEEEtran}
\bibliography{IEEEabrv,references}

% Generated by IEEEtran.bst, version: 1.14 (2015/08/26)
\begin{thebibliography}{10}
\providecommand{\url}[1]{#1}
\csname url@samestyle\endcsname
\providecommand{\newblock}{\relax}
\providecommand{\bibinfo}[2]{#2}
\providecommand{\BIBentrySTDinterwordspacing}{\spaceskip=0pt\relax}
\providecommand{\BIBentryALTinterwordstretchfactor}{4}
\providecommand{\BIBentryALTinterwordspacing}{\spaceskip=\fontdimen2\font plus
\BIBentryALTinterwordstretchfactor\fontdimen3\font minus
  \fontdimen4\font\relax}
\providecommand{\BIBforeignlanguage}[2]{{%
\expandafter\ifx\csname l@#1\endcsname\relax
\typeout{** WARNING: IEEEtran.bst: No hyphenation pattern has been}%
\typeout{** loaded for the language `#1'. Using the pattern for}%
\typeout{** the default language instead.}%
\else
\language=\csname l@#1\endcsname
\fi
#2}}
\providecommand{\BIBdecl}{\relax}
\BIBdecl

\bibitem{Cheng2024100Gbps}
C.~Chen, S.~Das, S.~Videv, A.~Sparks, S.~Babadi, A.~Krishnamoorthy, C.~Lee,
  D.~Grieder, K.~Hartnett, P.~Rudy, J.~Raring, M.~Najafi, V.~K. Papanikolaou,
  R.~Schober, and H.~Haas, ``100 {Gbps} indoor access and 4.8 {Gbps} outdoor
  point-to-point {LiFi} transmission systems using laser-based light sources,''
  \emph{J. Lightwave Technol.}, vol.~42, pp. 4146--4157, Jun. 2024.

\bibitem{THZ_compare}
M.~Shehata, Y.~Wang, J.~He, S.~Kandeepan, and K.~Wang, ``Optical and terahertz
  wireless technologies: {The} race to {6G} communications,'' \emph{IEEE Wirel.
  Commun.}, vol.~30, no.~5, pp. 10--18, 2023.

\bibitem{Liu2024}
M.~Liu, H.~Kazemi, M.~Safari, I.~Tavakkolnia, and H.~Haas, ``Energy efficiency
  comparison of {THz} and {VCSEL-}based {OWC} for {6G},'' Apollo - University
  of Cambridge Repository, Oct. 2024.

\bibitem{Soltani2023}
M.~D. Soltani, A.~A. Qidan, S.~Huang, B.~Yosuf, S.~Mohamed, R.~Singh, Y.~Liu,
  W.~Ali, R.~Chen, H.~Kazemi, E.~Sarbazi, B.~Berde, D.~Chiaroni,
  B.~Béchadergue, F.~Abdel-dayem, H.~Soni, J.~Tabu, M.~Perrufel,
  N.~Serafimovski, T.~E. El-Gorashi, J.~Elmirghani, M.~Crisp, R.~Penty, I.~H.
  White, H.~Haas, and M.~Safari, ``Terabit indoor laser-based wireless
  communications: {LiFi} $2.0$ for {6G},'' \emph{IEEE Wirel. Commun.}, vol.~30,
  pp. 36--43, Oct. 2023.

\bibitem{THZ_350GHz}
C.~Wang, B.~Lu, C.~Lin, Q.~Chen, L.~Miao, X.~Deng, and J.~Zhang, ``0.34-{THz}
  wireless link based on high-order modulation for future wireless local area
  network applications,'' \emph{IEEE Trans. Terahertz Sci. Technol.}, vol.~4,
  no.~1, pp. 75--85, 2014.

\bibitem{Giordani2020toward6G}
M.~Giordani, M.~Polese, M.~Mezzavilla, S.~Rangan, and M.~Zorzi, ``{Toward 6G
  Networks: Use Cases and Technologies},'' \emph{IEEE Communications Magazine},
  vol.~58, no.~3, pp. 55--61, 2020.

\bibitem{Koonen2020ultra}
T.~Koonen, K.~Mekonnen \emph{et~al.}, ``{Ultra-high-capacity wireless
  communication by means of steered narrow optical beams},''
  \emph{{Philosophical Transactions of the Royal Society A: Mathematical,
  Physical and Engineering Sciences}}, vol. 378, 2020.

\bibitem{Sarbazi2020TOWS_AP}
E.~Sarbazi, H.~Kazemi, M.~D. Soltani, M.~Safari, and H.~Haas, ``{A Tb/s Indoor
  Optical Wireless Access System Using VCSEL Arrays},'' in \emph{Proc. IEEE
  31st Annual International Symposium on Personal, Indoor and Mobile Radio
  Communications (PIMRC)}, 2020, pp. 1--6.

\bibitem{Kazemi2024TOWS_GoB}
H.~Kazemi, E.~Sarbazi, M.~Crisp, T.~E.~H. El-Gorashi, J.~M.~H. Elmirghani,
  R.~V. Penty, I.~H. White, M.~Safari, and H.~Haas, ``{A Novel Terabit
  Grid-of-Beam Optical Wireless Multi-User Access Network With Beam
  Clustering},'' \emph{{arXiv Preprint arXiv:2404.04443 [eess.SP]}}, Apr. 2024.

\bibitem{Tezergil2022WirelessBackhaul}
B.~Tezergil and E.~Onur, ``Wireless backhaul in {5G} and beyond: {Issues},
  challenges and opportunities,'' \emph{IEEE Commun. Surveys \& Tut.}, vol.~24,
  pp. 2579--2632, Fourthquarter 2022.

\bibitem{Schulz2016RobustOptical}
D.~Schulz, V.~Jungnickel, C.~Alexakis, M.~Schlosser, J.~Hilt,
  A.~Paraskevopoulos, L.~Grobe, P.~Farkas, and R.~Freund, ``Robust optical
  wireless link for the backhaul and fronthaul of small radio cells,'' \emph{J.
  Lightwave Technol.}, vol.~34, pp. 1523--1532, Mar. 2016.

\bibitem{Kazemi2022ATbps}
H.~Kazemi, E.~Sarbazi, M.~D. Soltani, T.~E.~H. El-Gorashi, J.~M.~H. Elmirghani,
  R.~V. Penty, I.~H. White, M.~Safari, and H.~Haas, ``A {Tb/s} indoor {MIMO}
  optical wireless backhaul system using {VCSEL} arrays,'' \emph{IEEE Trans.
  Commun.}, vol.~70, pp. 3995--4012, Jun. 2022.

\bibitem{Sarbazi2024Receiver}
E.~Sarbazi, H.~Kazemi, M.~Crisp, T.~El-Gorashi, J.~Elmirghani, R.~V. Penty,
  I.~H. White, M.~Safari, and H.~Haas, ``{Design and Optimization of High-Speed
  Receivers for 6G Optical Wireless Networks},'' \emph{IEEE Transactions on
  Communications}, vol.~72, no.~2, pp. 971--990, 2024.

\bibitem{UWOC_AquaFi}
B.~Shihada, O.~Amin, C.~Bainbridge, S.~Jardak, O.~Alkhazragi, T.~K. Ng, B.~Ooi,
  M.~Berumen, and M.-S. Alouini, ``Aqua-fi: Delivering internet underwater
  using wireless optical networks,'' \emph{IEEE Communications Magazine},
  vol.~58, no.~5, pp. 84--89, Jun. 2020.

\bibitem{Celik2019DCN}
A.~Celik, B.~Shihada, and M.-S. Alouini, ``{Optical wireless data center
  networks: potentials, limitations, and prospects},'' in \emph{Broadband
  Access Communication Technologies XIII}, vol. 10945, International Society
  for Optics and Photonics.\hskip 1em plus 0.5em minus 0.4em\relax SPIE, 2019,
  p. 109450I.

\bibitem{Zhou2020}
H.~Zhou, W.~Xu, J.~Chen, and W.~Wang, ``Evolutionary {V2X} technologies toward
  the {Internet of Vehicles}: {Challenges} and opportunities,''
  \emph{Proceedings of the IEEE}, vol. 108, no.~2, pp. 308--323, February 2020.

\bibitem{kong2020survey}
M.~Kong, C.~H. Kang, O.~Alkhazragi, X.~Sun, Y.~Guo, M.~Sait, J.~A.
  Holguin-Lerma, T.~K. Ng, and B.~S. Ooi, ``Survey of energy-autonomous solar
  cell receivers for satellite--air--ground--ocean optical wireless
  communication,'' \emph{Progress in quantum electronics}, vol.~74, p. 100300,
  2020.

\bibitem{green2024solar}
M.~A. Green, E.~D. Dunlop, M.~Yoshita, N.~Kopidakis, K.~Bothe, G.~Siefer,
  D.~Hinken, M.~Rauer, J.~Hohl-Ebinger, and X.~Hao, ``Solar cell efficiency
  tables (version 64),'' \emph{Progress in photovoltaics: research and
  applications}, vol.~32, no.~7, pp. 425--441, 2024.

\bibitem{IRENA2024}
IRENA, ``Renewable power generation costs in $2023$,'' International Renewable
  Energy Agency (IRENA), Abu Dhabi, Tech. Rep., Sep. 2024.

\bibitem{helmers2024advancing}
H.~Helmers, O.~H{\"o}hn, D.~Lackner, P.~Schygulla, M.~Klitzke, J.~Sch{\"o}n,
  C.~Pellegrino, E.~Oliva, M.~Schachtner, P.~Beutel \emph{et~al.}, ``Advancing
  solar energy conversion efficiency to 47.6\% and exploring the spectral
  versatility of {III-V} photonic power converters,'' in \emph{Physics,
  Simulation, and Photonic Engineering of Photovoltaic Devices XIII}, vol.
  12881.\hskip 1em plus 0.5em minus 0.4em\relax SPIE, 2024, pp. 6--15.

\bibitem{hohn2016optimal}
O.~H{\"o}hn, A.~Walker, A.~Bett, and H.~Helmers, ``Optimal laser wavelength for
  efficient laser power converter operation over temperature,'' \emph{Applied
  Physics Letters}, vol. 108, no.~24, 2016.

\bibitem{hohn2022realization}
O.~H{\"o}hn, M.~Schauerte, P.~Schygulla, H.~Hauser, D.~Lackner, B.~Bl{\"a}si,
  and H.~Helmers, ``Realization of ultrathin {GaAs} photonic power converters
  with rear-side metal grating on full 4\textsf{''} wafers,'' in \emph{2022
  IEEE 49th Photovoltaics Specialists Conference (PVSC)}.\hskip 1em plus 0.5em
  minus 0.4em\relax IEEE, 2022, pp. 0038--0042.

\bibitem{6213058}
O.~D. Miller, E.~Yablonovitch, and S.~R. Kurtz, ``Strong internal and external
  luminescence as solar cells approach the {Shockley–Queisser} limit,''
  \emph{IEEE Journal of Photovoltaics}, vol.~2, no.~3, pp. 303--311, 2012.

\bibitem{helmers202168}
H.~Helmers, E.~Lopez, O.~H{\"o}hn, D.~Lackner, J.~Sch{\"o}n, M.~Schauerte,
  M.~Schachtner, F.~Dimroth, and A.~W. Bett, ``68.9\% efficient {GaAs}-based
  photonic power conversion enabled by photon recycling and optical
  resonance,'' \emph{physica status solidi (RRL)--Rapid Research Letters},
  vol.~15, no.~7, p. 2100113, 2021.

\bibitem{fakidis2020simultaneous}
J.~Fakidis, H.~Helmers, and H.~Haas, ``Simultaneous wireless data and power
  transfer for a {1-Gb/s} {GaAs} {VCSEL} and photovoltaic link,'' \emph{IEEE
  Photonics Technology Letters}, vol.~32, no.~19, pp. 1277--1280, 2020.

\bibitem{kimovec2019comprehensive}
R.~Kimovec, H.~Helmers, A.~W. Bett, and M.~Topi{\v{c}}, ``Comprehensive
  electrical loss analysis of monolithic interconnected multi-segment laser
  power converters,'' \emph{Progress in Photovoltaics: Research and
  Applications}, vol.~27, no.~3, pp. 199--209, 2019.

\bibitem{de2024reconfigurable}
J.~I. De~Oliveira~Filho, A.~Trichili, O.~Alkhazragi, M.-S. Alouini, B.~S. Ooi,
  and K.~N. Salama, ``Reconfigurable {MIMO}-based self-powered battery-less
  light communication system,'' \emph{Light: Science \& Applications}, vol.~13,
  no.~1, p. 218, 2024.

\bibitem{wang2012terabit}
J.~Wang, J.-Y. Yang, I.~M. Fazal, N.~Ahmed, Y.~Yan, H.~Huang, Y.~Ren, Y.~Yue,
  S.~Dolinar, M.~Tur \emph{et~al.}, ``Terabit free-space data transmission
  employing orbital angular momentum multiplexing,'' \emph{Nature photonics},
  vol.~6, no.~7, pp. 488--496, 2012.

\bibitem{willner2016design}
A.~E. Willner, G.~Xie, L.~Li, Y.~Ren, Y.~Yan, N.~Ahmed, Z.~Zhao, Z.~Wang,
  C.~Liu, A.~J. Willner \emph{et~al.}, ``Design challenges and guidelines for
  free-space optical communication links using orbital-angular-momentum
  multiplexing of multiple beams,'' \emph{Journal of Optics}, vol.~18, no.~7,
  p. 074014, 2016.

\bibitem{Najafi2019RIS}
M.~Najafi and R.~Schober, ``Intelligent reflecting surfaces for free space
  optical communications,'' in \emph{IEEE Global Communications Conference
  (GLOBECOM)}, Dec. 2019.

\bibitem{DigitalRIS}
A.~R. Ndjiongue, T.~M.~N. Ngatched, O.~A. Dobre, and H.~Haas, ``Digital {RIS
  (DRIS):} {The} future of digital beam management in {RIS}-assisted {OWC}
  systems,'' \emph{Journal of Lightwave Technology}, vol.~40, no.~16, pp.
  5597--5604, 2022.

\bibitem{maraqa2023optimized}
O.~Maraqa and T.~M. Ngatched, ``Optimized design of joint mirror array and
  liquid crystal-based {RIS}-aided {VLC} systems,'' \emph{IEEE Photonics
  Journal}, 2023.

\bibitem{huang2024energy}
Y.~Huang, V.~K. Papanikolaou, H.~Ajam, M.~Safari, R.~Schober, H.~Haas, and
  I.~Tavakkolnia, ``Energy-efficient {RIS}-aided laser-based {LiFi} system with
  dynamic coverage optimization,'' in \emph{Proceedings of the IEEE
  International Conference on Communications (ICC)}.\hskip 1em plus 0.5em minus
  0.4em\relax IEEE, 2024.

\bibitem{yoshida2020245}
K.~Yoshida, P.~P. Manousiadis, R.~Bian, Z.~Chen, C.~Murawski, M.~C. Gather,
  H.~Haas, G.~A. Turnbull, and I.~D. Samuel, ``245 {MHz} bandwidth organic
  light-emitting diodes used in a gigabit optical wireless data link,''
  \emph{Nature communications}, vol.~11, no.~1, p. 1171, 2020.

\bibitem{10308568}
M.~N. Munshi, L.~Maret, B.~Racine, A.~P.~A. Fischer, M.~Chakaroun, and
  N.~Loganathan, ``2.85-{Gb/s} organic light communication with a 459-{MHz}
  micro-{OLED},'' \emph{IEEE Photonics Technology Letters}, vol.~35, no.~24,
  pp. 1399--1402, 2023.

\bibitem{yoshida2024rgb}
K.~Yoshida, C.~Chen, H.~Haas, G.~A. Turnbull, and I.~D. Samuel,
  ``{RGB}-single-chip {OLEDs} for high-speed visible-light communication by
  wavelength-division multiplexing,'' \emph{Advanced Science}, p. 2404576,
  2024.

\bibitem{zhu2024high}
Y.~Zhu, H.~Chen, R.~Han, H.~Qin, Z.~Yao, H.~Liu, Y.~Ma, X.~Wan, G.~Li, and
  Y.~Chen, ``High-speed flexible near-infrared organic photodiode for optical
  communication,'' \emph{National Science Review}, vol.~11, no.~3, p. nwad311,
  2024.

\bibitem{liguori2024overcoming}
R.~Liguori, F.~Nunziata, S.~Aprano, and M.~G. Maglione, ``Overcoming challenges
  in {OLED} technology for lighting solutions,'' \emph{Electronics}, vol.~13,
  no.~7, p. 1299, 2024.

\bibitem{zhang2023organic}
X.~Zhang, J.~Jiang, B.~Feng, H.~Song, and L.~Shen, ``Organic photodetectors:
  {Materials}, device, and challenges,'' \emph{Journal of Materials Chemistry
  C}, vol.~11, no.~37, pp. 12\,453--12\,465, 2023.

\bibitem{Tian2022}
P.~Tian, X.~Shan, S.~Zhu, E.~Xie, J.~J.~D. McKendry, E.~Gu, and M.~D. Dawson,
  ``{AlGaN} ultraviolet micro-{LEDs},'' \emph{IEEE Journal of Quantum
  Electronics}, vol.~58, pp. 1--14, Mar. 2022.

\bibitem{Maclure2022a}
D.~M. Maclure, J.~J.~D. McKendry, M.~S. Islim, E.~Xie, C.~Chen, X.~Sun,
  X.~Liang, X.~Huang, H.~Abumarshoud, J.~Herrnsdorf, E.~Gu, H.~Haas, and M.~D.
  Dawson, ``10 {Gbps} wavelength division multiplexing using {UV-A, UV-B, and
  UV-C} micro-{LEDs},'' \emph{Photon. Res.}, vol.~10, no.~2, pp. 516--523, Feb
  2022.

\bibitem{Maclure2022b}
D.~M. Maclure, C.~Chen, J.~J.~D. McKendry, E.~Xie, J.~Hill, J.~Herrnsdorf,
  E.~Gu, H.~Haas, and M.~D. Dawson, ``Hundred-meter {Gb/s} deep ultraviolet
  wireless communications using {AlGaN} micro-{LEDs},'' \emph{Opt. Express},
  vol.~30, no.~26, pp. 46\,811--46\,821, Dec 2022.

\bibitem{zimi2023gigabit}
H.~Zimi, D.~Maclure, C.~Chen, J.~McKendry, D.~Stothard, J.~Herrnsdorf, H.~Haas,
  and M.~Dawson, ``{Gigabit per second UV-C LEDs for communications},'' in
  \emph{2023 IEEE Photonics Conference (IPC)}.\hskip 1em plus 0.5em minus
  0.4em\relax IEEE, 2023, pp. 1--2.

\bibitem{Nicholls2023}
J.~Nicholls, L.~Anderson, W.~Lee, J.~J.~S. Ahn, A.~Baskaran, H.~Bang,
  M.~Belloeil, Y.~Cai, J.~Campbell, J.~Chai, N.~Corpuz, V.~Entoma, B.~Hayden,
  T.~Hung, H.~Kim, D.~King, S.~Li, A.~Liu, D.~McMahon, V.~Nguyen, S.~F. Pan,
  S.~Tedman-Jones, W.~J. Toe, R.~Tsai, M.~P. Tudo, H.~P. Wang, Y.~Wang, S.~Yan,
  R.~Yang, K.~Yeo, W.~Schaff, N.~Krause, R.~Charters, J.~Tang, and
  P.~Atanackovic, ``High performance and high yield sub-240 nm {AlN:GaN} short
  period superlattice {LEDs} grown by {MBE} on 6 in. sapphire substrates,''
  \emph{Appl. Phys. Lett.}, vol. 123, p. 051105, Aug. 2023.

\bibitem{bakar2020accurate}
A.~H.~A. Bakar, T.~Glass, H.~Y. Tee, F.~Alam, and M.~Legg, ``Accurate visible
  light positioning using multiple-photodiode receiver and machine learning,''
  \emph{IEEE Trans. Instrum. Meas.}, vol.~70, pp. 1--12, Sep. 2020.

\bibitem{he2019demonstration}
J.~He, C.-W. Hsu, Q.~Zhou, M.~Tang, S.~Fu, D.~Liu, L.~Deng, and G.-K. Chang,
  ``{Demonstration of high precision 3D indoor positioning system based on
  two-layer ANN machine learning technique},'' in \emph{Optical Fiber Commun.
  Conf. and Ex. (OFC)}.\hskip 1em plus 0.5em minus 0.4em\relax IEEE, 2019, pp.
  1--3.

\bibitem{ahmad2023joint}
R.~Ahmad, H.~Kazemi, E.~Sarbazi, and H.~Haas, ``{Joint Position and Orientation
  Estimation in VCSEL-Based LiFi Networks: A Deep Learning Approach},'' in
  \emph{IEEE Global Commun. Conf. (GLOBECOM)}.\hskip 1em plus 0.5em minus
  0.4em\relax IEEE, 2023, pp. 3676--3681.

\bibitem{kokdogan2024intelligent}
F.~Kokdogan and S.~Gezici, ``Intelligent reflecting surfaces for visible light
  positioning based on received power measurements,'' \emph{IEEE Trans. Veh.
  Technol.}, Apr. 2024.

\bibitem{Gesture3}
Y.~Li, T.~Li, R.~A. Patel, X.-D. Yang, and X.~Zhou, ``Self-powered gesture
  recognition with ambient light,'' in \emph{Proc. 31st Annual ACM Symp. on
  User Interf. Softw. and Tech.}, ser. UIST '18.\hskip 1em plus 0.5em minus
  0.4em\relax New York, NY, USA: Association for Computing Machinery, 2018, p.
  595–608.

\bibitem{Gesture1}
X.~Liang, J.~Li, C.~Xu, Z.~Xie, C.~Zhang, W.~Ding, and W.~Gui, ``Gesture
  recognition and control based on visible light communication,'' in
  \emph{Chinese Control Conf. (CCC)}, 2024, pp. 6164--6169.

\end{thebibliography}

\end{document}